\documentclass[twocolumn,preprintnumbers,amssymb,aps]{revtex4-1}
\usepackage{pgfplots}
\usepackage{latexsym,epsfig}
\usepackage{graphicx}
\usepackage{ulem}
\usepackage{dcolumn}
\usepackage{color}
\usepackage{amsthm,amsmath}

\begin{document}

\title{Constraints on new physics from an improved calculation of parity violation in $^{133}$Cs}

\author{B. K. Sahoo}
\email{bijaya@prl.res.in}
\affiliation{Atomic, Molecular and Optical Physics Division, Physical Research Laboratory, Navrangpura, Ahmedabad-380009, India}
\author{B. P. Das}
\affiliation{Department of Physics, School of Science, Tokyo Institute of Technology,
2-1-2-1-H86 Ookayama Meguro-ku, Tokyo 152-8550, Japan \\
Centre for Quantum Engineering Research and Education, TCG Centres for Research in Science and Technology, Sector V, Salt Lake, Kolkata 70091, India}

\begin{abstract}
We report the result of our calculation of the nuclear spin-independent parity violating electric dipole transition amplitude ($E1_{PV}$) for the 
$6s ~ ^2S_{1/2} - 7s ~ ^2S_{1/2}$ transition in $^{133}$Cs to an accuracy of 0.3\% using a variant of the perturbed relativistic coupled-cluster (RCC)
theory. In the present work, we treat the contributions of both the low-lying and high-lying excited states to the above mentioned amplitude on the 
same footing, thereby overcoming the limitations of previous high accuracy RCC calculations. We obtain an accurate value for the vector polarizability
($\beta$) for the above transition and by combining it with the results from our present calculation of $E1_{PV}$  and the latest measurement of 
$Im(E1_{PV}/\beta)$, we extract the nuclear weak charge ($Q_W$); and analyze its deviation from its value in the Standard Model (SM) in
order to constrain certain scenarios of new physics beyond the SM.
\end{abstract}

\date{Received date; Accepted date}

\maketitle

\section{Introduction}

Atomic processes are usually studied by considering the exchange of photons ($\gamma$) between the bound electrons and the nucleus and the bound 
electrons themselves. The longitudinal $\gamma$s are responsible for the Coulomb interaction, which is the dominant  contribution to electromagnetic 
interactions in atomic systems. However, the Breit interaction \cite{Breit} due to the transverse $\gamma$s and quantum electrodynamics (QED) effects 
must also be considered in high precision atomic calculations. All these interactions preserve parity symmetry, and these systems are described 
conveniently using spherical coordinates \cite{Griffiths}. Inclusion of the neutral current weak interactions due to the exchange of $Z_0$ boson in 
atomic systems leads to parity violation \cite{Bouchiat}, and this phenomenon has been referred to as atomic parity violation (APV). Depending on the 
electron-nucleus vector--axial-vector (V-A) or the axial-vector--vector (A-V) currents, APV interactions can be nuclear spin ($I$) independent or
dependent. In addition to the $Z_0$ exchange interactions, the possible interaction of the nuclear anapole moment with electrons can give rise to 
APV that depends on $I$ \cite{Ginges1}. The $I$ dependent APV contributions are relatively smaller than its NSI counterpart, as the odd-nucleon 
contributes primarily to APV. In the case of the nuclear spin independent (NSI) interaction, the nucleons which ultimately arise from the up- and 
down-quarks, contribute coherently, and these contributions are several orders of magnitude larger than those due to the exchange of $Z_0$ between 
the electrons \cite{Bouchiat,Ginges1,Cahn,Marciano00}. Nonetheless, the NSI APV interaction is too weak to be detected using typical spectroscopic
measurements, and therefore, special techniques have been developed in different laboratories to observe effects due to it 
\cite{Commins,Meekhof,wood,bouchiat1,Tsigutkin}. Heavier atomic systems are preferred for measuring the subtle NSI APV effects owing to the fact that 
the weak interaction causing them in atomic systems scales slightly faster than $Z^3$ \cite{Bouchiat}, where $Z$ is the atomic number. These 
measurements in combination with high-precision atomic calculations have the potential to probe physics beyond the Standard Model (SM) of 
elementary particles \cite{marciano,erler,kumar,Cirigliano,derevianko1}.

APV has been measured to an accuracy of 0.35\% in the $6s ~ ^2S_{1/2} - 7s ~ ^2S_{1/2}$ transition in $^{133}$Cs \cite{wood}. This is the most 
accurate APV measurement to date. New experiments have been proposed to measure APV in Cs \cite{choi,anders}, which have the 
potential to surpass the accuracy of this measurement. The stage is now clearly set to take the APV calculations in Cs to the next level, 
which would lead to an improvement in the accuracy of $Q_W$, thereby making it possible to probe new physics beyond the SM. This indeed provides 
the motivation for our present work. The principal quantity of interest in the APV studies is the nuclear weak charge ($Q_W$), which is a linear 
combination of the coupling coefficients between electrons, and up- and down-quarks in an atomic system \cite{Bouchiat,bouchiat1}. The difference in 
the model independent value of $Q_W$ obtained from APV and that obtained from the SM could, in principle, shed light on new physics beyond the SM. 
The APV study in $^{133}$Cs currently yields $\sin^2 \theta_W^{\rm exp}=0.2356(20)$ \cite{dzuba1} and the value predicted by the SM is $\sin^2 
\theta_W^{\rm SM}=0.23857(5)$ \cite{tanabashi}, for the Weinberg angle $\theta_W$, at the zero momentum transfer in the $\overline{\rm MS}$ scheme. 
The difference in the values of  $Q_W$ also gives the lower mass limit of an extra $Z_{\chi}$ boson as $M_{Z_{\chi}} > 710$ GeV$/c^2$ and weak isospin
conserving parameter $S=-0.81(54)$ at the 1$\sigma$ level \cite{dzuba1}. It has been argued that the signature of a dark boson ($Z_d$) (also referred 
to as dark photon in the literature) can be obtained from the above mentioned difference \cite{hooman,marciano1,davoudias}. Direct experimental 
signature suggests its value to be less than 0.2 GeV \cite{Banerjee}. A fairly recent manuscript \cite{arcadi} suggests that the limit on an 
effective electron-nucleus coupling describing new physics beyond the SM expressed as $\frac{f_{\rm Vq}^{eff}}{\Lambda^2} < 4.38699 \times 10^{-9}$
GeV$^{-2}$ with a new energy scale $\Lambda$ and emphasizes that the bound on the effective couplings inferred from APV is more stringent than 
the ones from the neutrino-nucleus coherent scattering processes.

 State-of-the art relativistic atomic many-body theories have been applied to $^{133}$Cs APV calculations in the last four decades \cite{dzuba1,
dzuba2,blundell,Derevianko3,kozlov,shabaev,porsev1}. The accuracy of the calculations have steadily improved, as the theoretical methods have been
able to incorporate larger classes of higher-order effects during this time due to advances in high performance computing. The latest two high
precision calculations have been reported in Refs. \cite{dzuba1,porsev1}. These calculations divided the entire electron correlation contribution 
into three parts and they are calculated by mixed many-body methods. Further, the dominant part was evaluated through sum-over-states approach and 
the other contributions were not treated on the same footing. Contributions from the Breit and QED effects were taken from the earlier works, but not 
double core-polarization (DCP) effects \cite{roberts2}. Some of these issues triggered discussions recently \cite{derevianko1,derevianko2}, and 
therefore, it is necessary to revisit this problem. In this work, we intend to circumvent the above mentioned  limitations of the previous 
calculations by solving the first-order perturbed wave functions due to the APV interaction for atomic states in the framework of the relativistic 
coupled-cluster (RCC) theory. Thus, it considers both the electromagnetic and weak interactions simultaneously in addition to accounting for Coulomb, 
Breit and QED interactions using the same many-body method. Most importantly, it treats all the three different parts of the total correlation 
contribution mentioned above on the same footing.

\section{Theory}

Neglecting the A-V interaction, the short range effective Lagrangian corresponding to the V-A neutral weak current interaction of an electron 
with up- and down-quarks in an atomic system is given by \cite{Marciano00,Commins}
\begin{eqnarray}
 {\cal L}_{\rm eq}^{\rm VA} &=& \frac{G_F}{\sqrt{2}} \sum_{u,d} \left [ C_{1u} \bar{\psi}_{u} \gamma_{\mu} \psi_{u} + 
 C_{1d} \bar{\psi}_{d} \gamma^{\mu} \psi_{d} \right ] \bar{\psi}_e \gamma^{\mu} \gamma^5 \psi_e \nonumber \\
 &=& \frac{G_F}{\sqrt{2}}  \sum_n C_{1n} \bar{\psi}_{n} \gamma_{\mu} \psi_{n} \bar{\psi}_e \gamma^{\mu} \gamma^5 \psi_e ,
\end{eqnarray}
where $G_F=1.16632 \times 10^{-5}$ GeV$^{-2}$ is the Fermi constant, sums $u$, $d$ and $n$ stand for up-quark, down-quark and nucleons 
respectively, and $C_{i=u,d,n}$ represent coupling coefficients of the interaction of an electron with quarks and nucleons. Adding them 
coherently and taking the non-relativistic approximation for nucleons, the temporal component gives the NSI weak interaction Hamiltonian
\begin{eqnarray}
 H_{\rm en}^{\rm VA} &=& -\frac{G_F}{2\sqrt{2}} \left [ Q_W \rho(r) + (N C_{1N} - Z C_{1P}) \Delta \rho(r) \right ] \gamma^5 , \ \ \
\end{eqnarray}
where $N$ and $P$ representing for neutron and proton respectively, $\rho(r)=(\rho_N(r)+\rho_P(r))/2$ is the average nucleon density with 
normalized proton density $\rho_P(r)$ and normalized neutron density $\rho_N(r)$, $\Delta \rho(r)=\rho_N(r) - \rho_P(r)$, and 
$Q_W=2[ZC_{1P}+NC_{1P}]$ is known as the nuclear weak charge. In the atomic calculations, contribution from $\Delta \rho(r)$ is neglected 
at first, but is added later as ``nuclear skin'' correction. The nuclear skin correction to $Q_W$ is expressed as \cite{sil}
\begin{eqnarray}
 \Delta Q_W^{N-P} = 0.9857 N \frac{(Z\alpha_e)^2}{q_P} \frac{232}{525}\frac{t}{r_P} ,
\end{eqnarray}
where $\alpha_e$ is the fine-structure constant, $r_{i=P(N)}$ are the root mean square radius of proton (neutron), $t=r_N-r_P$ is the neutron skin, and 
$q_P$ is defined as
\begin{eqnarray}
 q_P = \int d^3r f(r) \rho_P(r) 
\end{eqnarray}
with the electronic form factor $f(r)$ that describes the spatial variation of the electronic axial-vector matrix element over the size of the 
nucleus.

The NSI weak interaction Hamiltonian for atomic calculations, thus, is given by \cite{Bouchiat}
\begin{eqnarray}
\label{eq1}
H_{APV}^{NSI}&=& \sum_e H_{\rm en}^{\rm AV}=-\frac{G_F}{2\sqrt{2}}Q_{W}^{\rm at} \sum_e \gamma_e^{5} \rho(r_e) ,
\end{eqnarray}
where $Q_W^{\rm at} = Q_W - \Delta Q_W^{N-P}$. It is obvious that $Q_W$ is a model dependent quantity. Thus, the difference of its actual value 
from the SM, given by $\Delta Q_W = Q_W^{\rm exp} -Q_W^{\rm SM}$, can provide signatures about new physics. In the SM, $C_{1u}=\frac{1}{2}\left
[1- \frac{8}{3}\sin^2 \theta_W^{\rm SM} \right ]$ and $C_{1d}=-\frac{1}{2}\left [1- \frac{4}{3}\sin^2 \theta_W^{\rm SM} \right ]$ \cite{Cahn,
Marciano00,Commins,bouchiat1}. This follows $C_{1N}=2C_{1d}+C_{1u} = -1/2$ and $C_{1P}=2C_{1u}+C_{1d}=(1-4 \sin^2 \theta_W^{\rm SM})/2 \approx 0.04$. 

Moreover, $\sin^2 \theta_W$ varies with energy scale (denoted by $Q$) and is parameterized in the $\overline{\rm MS}$ scheme as \cite{kumar}
\begin{eqnarray}
 \sin^2 \theta_W(Q^2) = \kappa(Q^2) \sin^2 \theta_W(M_{Z_0})_{\overline{\rm MS}}, 
\end{eqnarray}
where $M_{Z_0}$ is the mass of $Z_0$-boson and $\kappa(Q^2)$ denotes perturbative $\gamma$ -- $Z_0$-boson mixing. For the normalization $\kappa(Q^2 
\equiv M_{Z_0}^2)=1.0$, it corresponds to $\kappa(0) \sim 1.03$ \cite{kumar}. In the one-loop radiative correction, the mass of $W$-boson 
and $\sin^2 \theta_W(m_{Z_0})_{\overline{\rm MS}}$ are given by \cite{kumar}
\begin{eqnarray}
 M_W = 80.362(6) [1-0.0036S+0.0056T] \ \  {\rm GeV}/c^2 
 \end{eqnarray}
 and
\begin{eqnarray}
 \sin^2 \theta_W(m_{Z_0})_{\overline{\rm MS}} = 0.23124(6) [1 + 0.0157S - 0.0112 T], \ \ \ \
\end{eqnarray}
where $c$ is speed of light, and $S$ and $T$ are the isospin conserving and isospin breaking parameters, respectively. By comparing the above 
expression for $M_W$ with its experimental value of $80.379(12)$ GeV/$c^2$ \cite{tanabashi}, it gives \cite{marciano1} 
\begin{eqnarray}
 S = 0.07 \pm 0.09 \ \ \ \ {\rm and} \ \ \ \ T = 0.10 \pm 0.09 .
\end{eqnarray}
Also, $M_{Z_{\chi}}$ can be obtained in the SO(10) model as \cite{marciano}
\begin{eqnarray}
\Delta Q_W \approx 0.4 \times (2N+Z) (M_W/M_{Z_{\chi}})^2 .
\end{eqnarray}

In the $Z_0$ and $Z_d$ mixing of two-Higgs doublet model scenario, we get \cite{marciano,marciano1,davoudias}
\begin{eqnarray}
 \sin^2 \theta_W^{\rm exp}(0) - \sin^2 \theta_W^{\rm SM}(0) \simeq -0.42 \varepsilon \delta \frac{M_{Z_0}}{M_{Z_d}} ,
\end{eqnarray}
where $\varepsilon$ and $\delta$ are the model dependent parameters, and $M_{Z_d}$ is mass of $Z_d$.

Accounting for all the aforementioned possible physics beyond the SM, the weak charge of $^{133}$Cs atom can be expressed 
in terms of all the combined parameters as
\begin{eqnarray}
 Q_W({\rm ^{133}Cs}) &=& Q_W^{\rm SM}({\rm ^{133}Cs}) \times [ 1+0.011 S -0.008T \nonumber \\ 
 &-& 0.9 (M_{Z_0}^2 / M_{Z_\chi}^2) - 1.265 \varepsilon \delta (M_{Z_0}/ M_{Z_d})] , \ \ \ \ \
\end{eqnarray}
where $Q_W^{\rm SM}({\rm ^{133}Cs})=-N+Z(1-4\sin^2 \theta_W^{\rm SM})=-73.23(1)$ is the nuclear weak charge of $^{133}$Cs in the SM \cite{tanabashi}. 

In an effective description \cite{arcadi}, $\Delta Q_W$  is encoded using a new energy scale $\Lambda$ as
\begin{eqnarray}
 \Delta Q_W = \frac{2\sqrt{2}}{G_F}\frac{3}{\Lambda^2} f_{Vq}^{\rm eff} (Z+N) ,
\end{eqnarray}
where 
\begin{eqnarray}
 f_{Vq}^{\rm eff} = \frac{C_{1u}(2Z+N)+C_{1d}(Z+2N)}{3(Z+N)} .
\end{eqnarray}

Similarly, $Q_W$ can be expressed in terms of the nucleon-electron V-A couplings as \cite{tanabashi}
\begin{eqnarray}
 Q_W = -2 \left [Zg^{ep} + Ng^{en}+0.00743 \right ] \left ( 1- \frac{\alpha_e}{\pi} \right ),
\end{eqnarray}
where $g^{ep(n)}$ are the electron-proton(neutron) coupling constants. The SM offers $55g^{ep}+78g^{en}=36.70(1)$ \cite{tanabashi}. 

\section{Atomic Calculations}

\subsection{General aspects}

The atomic wave function ($|\Psi_v \rangle$) of a state in Cs atom is calculated by dividing the total Hamiltonian as 
\begin{eqnarray}
 H = H_{em} + \lambda H_{w}, 
\label{eqham}
\end{eqnarray}
where $H_{em}$ represents the dominant electromagnetic interactions in the atom and $H_{APV}^{NSI} \equiv \lambda H_{w}$ with 
$\lambda=\frac{G_F}{2\sqrt{2}}Q_W^{\rm at}$. The electric dipole transition amplitude between the same nominal parity states $|\Psi_i\rangle$ and 
$|\Psi_f\rangle$ states due to the presence of $H_{APV}^{NSI}$ can be written as
\begin{eqnarray}
 E1_{PV} = \frac{\langle \Psi_f | D | \Psi_i \rangle}{\sqrt{\langle \Psi_f | \Psi_f \rangle \langle \Psi_i | \Psi_i \rangle}}.
\end{eqnarray}
Since the strength of $H_{APV}^{NSI}$ is much weaker than that of the $H_{em}$ in an atomic system, the wave function for a state 
(say, $|\Psi_v \rangle$) corresponding to the total Hamiltonian $H=H_{em}+\lambda H_{w}$ and its energy (say, $E_v$) can be expressed as
\begin{eqnarray}
 && |\Psi_v \rangle = |\Psi_v^{(0)} \rangle + \lambda |\Psi_v^{(1)} \rangle + {\cal O}(\lambda^2) \\
 \text{and} && \nonumber \\
  && E_v = E_v^{(0)} + \lambda E_v^{(1)}  + {\cal O}(\lambda^2) ,
\end{eqnarray}
where the superscripts 0 and 1 stand for the zeroth-order and first-order contributions due to $H_{w}$, respectively. By neglecting 
${\cal O}(\lambda^2)$ contributions, we get
\begin{eqnarray}
  E1_{PV} \simeq \lambda \frac{\langle \Psi_f^{(1)} | D | \Psi_i^{(0)} \rangle + \langle \Psi_f^{(0)} | D | \Psi_i^{(1)} \rangle}{\sqrt{\langle \Psi_f^{(0)} | \Psi_f^{(0)} \rangle \langle \Psi_i^{(0)} | \Psi_i^{(0)} \rangle}}.
\label{e1pnc}
\end{eqnarray}
As mentioned before, the previous two high-precision calculations of $E1_{PV}$ were evaluated using the sum-over-states approach by expanding the 
first-order wave function as
\begin{eqnarray}
 |\Psi_v^{(1)} \rangle = \sum_{I \ne v} |\Psi_I^{(0)} \rangle \frac{\langle \Psi_I^{(0)} | H_{w} | \Psi_v^{(0)} \rangle }{E_v^{(0)} - E_I^{(0)}} ,
\label{intst}
 \end{eqnarray}
where $I$ denotes all possible intermediate states, that can be divided into core states (contributions from these states are designated as ``Core''),
low-lying bound states (contributions from these states are given as ``Main''), and the remaining high-lying states including continuum (whose 
contributions are mentioned as ``Tail'') for computational simplicity. The drawback of this approach is that in an actual calculation, 
it is possible to evaluate ``Main'' contributions from only a few low-lying valence excited bound states accurately by calculating them individually 
using a powerful many-body method, and the ``Core'' and ``Tail'' contributions are estimated using less rigorous many-body methods. Therefore,
the results from the latter two sectors are less accurate. In other words, this approach of evaluating correlation effects in a piecemeal manner 
does not take into account certain types of correlation effects. As a consequence, contributions from 
effects such as the DCP  are completely excluded. Keeping in mind the high accuracy needed for APV to achieve its ultimate objective of probing
new physics beyond the SM, it is desirable to include contributions from all the intermediate states on an equal footing.
This can be accomplished not by summing over intermediate states, but rather by obtaining the first-order perturbed wave functions for the initial 
and final states directly. 

From the equation $H|\Psi_v \rangle = E_v |\Psi_v\rangle$, the inhomogeneous equation for the first-order wave function is obtained as 
\begin{eqnarray}
 (H_{em}-E_v^{(0)}) |\Psi_v^{(1)} \rangle = (E_v^{(1)}-H_{w})|\Psi_v^{(0)} \rangle ,
 \label{eqpt}
\end{eqnarray}
where $E_v^{(1)}=0$ in the present case owing to the odd-parity nature of $H_{w}$. Obtaining $|\Psi_v^{(1)} \rangle$ directly by solving the 
above equation can implicitly include contributions from all the intermediate states $I$ of Eq. (\ref{intst}), thereby, overcoming the problem
of unequal treatment of various electron correlation effects from different sectors as mentioned above. Moreover, it is also necessary to account for correlation effects involving both the weak and electromagnetic interactions. Therefore, it is very important to consider a powerful and versatile many-body theory to obtain both $|\Psi_v^{(0)} \rangle$ and 
$|\Psi_v^{(1)} \rangle$ accurately. Since Cs is a heavy atom, it is necessary to employ a relativistic method for computing the wave 
functions of this atom. The coupled-cluster (CC) theory is currently considered to be one of the leading quantum many-body methods and has been 
referred to as the gold standard for treating electron correlation effects in atomic and molecular systems \cite{bishop,bartlett,crawford}. Thus, the 
relativistic version of the CC (RCC) theory is very well suited for  the accurate evaluation of the correlation effects in  $E1_{PV}$ for the 
$6s ~ ^2S_{1/2} - 7s ~ ^2S_{1/2}$ transition in $^{133}$Cs. 

\subsection{Atomic Hamiltonian}

The starting point of our calculation is the Dirac-Coulomb (DC) Hamiltonian \cite{Dirac} representing the leading order contributions to $H_{em}$ to calculate
the zeroth-order wave functions and energies which in atomic units (a.u.) is given by
\begin{eqnarray}\label{eq:DHB}
H^{DC} &=& \sum_i \left [c\mbox{\boldmath$\alpha$}_i\cdot \textbf{p}_i+(\beta_i-1)c^2+V_n(r_i)\right] +\sum_{i,j>i}\frac{1}{r_{ij}}, \ \ \ \
\end{eqnarray}
where $\mbox{\boldmath$\alpha$}$ and $\beta$ are the usual Dirac matrices, $\textbf{p}$ is the single particle momentum operator, $V_n(r)$ 
denotes the nuclear potential, and $\sum_{i,j}\frac{1}{r_{ij}}$ represents the Coulomb potential between the electrons located at the 
$i^{th}$ and $j^{th}$ positions. It should be noted that the above Hamiltonian is scaled with respect to the rest mass energies of electrons.
Contributions from the Breit interaction \cite{breit} to $H_{em}$ is determined by including the following potential 
\begin{eqnarray}\label{eq:DHB}
V^B &=& - \sum_{j>i}\frac{[\mbox{\boldmath$\alpha$}_i\cdot\mbox{\boldmath$\alpha$}_j+
(\mbox{\boldmath$\alpha$}_i\cdot\mathbf{\hat{r}_{ij}})(\mbox{\boldmath$\alpha$}_j\cdot\mathbf{\hat{r}_{ij}})]}{2r_{ij}} ,
\end{eqnarray}
where $\mathbf{\hat{r}_{ij}}$ is the unit vector along $\mathbf{r_{ij}}$.

Contributions from the QED effects to $H_{em}$ are estimated by considering the lower-order vacuum polarization (VP) interaction ($V_{VP}$) 
and the self-energy (SE) interactions ($V_{SE}$). We account for $V_{VP}$ through the Uehling \cite{Uehl} 
and Wichmann-Kroll \cite{Wichmann} potentials ($V_{VP}=V^{Uehl} + V^{WK}$), given by
\begin{eqnarray}
 \label{eq:uehl}
V^{Uehl}&=&- \frac{2}{3} \sum_i \frac{\alpha_e^2 }{r_i} \int_0^{\infty} dx \ x \ \rho(x)\int_1^{\infty}dt \sqrt{t^2-1} \nonumber \\
&& \times\left(\frac{1}{t^3}+\frac{1}{2t^5}\right)  \left [ e^{-2ct|r_i-x|} - e^{-2ct(r_i+x)} \right ]\ \ 
\end{eqnarray}
and
\begin{eqnarray}
 V^{WK} = \sum_i \frac{0.368 Z^2}{9 \pi c^3 (1+(1.62 c r_i )^4) } \rho(r_i),
\end{eqnarray}
respectively.

The SE contribution $V_{SE}$ is estimated by including two parts \cite{Flambaum}
\begin{eqnarray}
V_{SE}^{ef}&=&  A_l \sum_i \frac{2 \pi Z \alpha_e^3 }{r_i} I_1^{ef}(r_i) - B_l \sum_i \frac{\alpha_e }{ r_i} I_2^{ef}(r_i) \ \ \
\end{eqnarray}
known as the effective electric form factor part and
\begin{eqnarray}
V_{SE}^{mg}&=& - \sum_k \frac{i\alpha_e^3}{4} \mbox{\boldmath$\gamma$} \cdot \mbox{\boldmath$\nabla$}_k \frac{1}{r_k} \int_0^{\infty} dx \ x \ \rho_n(x)
\int_1^{\infty} dt \frac{1}{t^3 \sqrt{t^2-1}} \nonumber \\
\times && \left [ e^{-2ct|r_k-x|} - e^{-2ct(r_k+x)} - 2ct \left (r_k+x-|r_k-x| \right ) \right ], \nonumber \\
\end{eqnarray}
known as the effective magnetic form factor part. In the above expressions, we use \cite{Ginges} 
\begin{eqnarray}
A_l= \begin{cases} 0.074+0.35Z \alpha_e \ \text{for} \ l=0,1 \\  0.056+0.05 Z \alpha_e + 0.195 Z^2 \alpha_e^2 \ \text{for} \ l=2  , \end{cases}
\end{eqnarray}
and
\begin{eqnarray}
B_l = \begin{cases} 1.071-1.97x^2 -2.128 x^3+0.169 x^4  \ \text{for} \ l=0,1 \\
     0 \ \text{for} \ l \ge 2 .  \end{cases}   
\end{eqnarray}
The integrals are given by
\begin{eqnarray}
I_1^{ef}(r) =  \int_0^{\infty} dx \ x \ \rho_n(x) [ (Z |r-x|+1) e^{-Z|r-x|} \nonumber \\  - (Z(r+x)+1) e^{-2ct(r+x)}  ] \ \ \ \ \ \ 
\end{eqnarray}
and
\begin{eqnarray}
 I_2^{ef}(r) &=& \int_0^{\infty} dx \ x \ \rho_n(x)  \int^{\infty}_1 dt \frac{1}{\sqrt{t^2-1}} \bigg \{ \left( 1-\frac{1}{2t^2} \right ) \nonumber \\
&\times& \left [ \ln(t^2-1)+4 \ln \left ( \frac{1}{Z \alpha_e} +\frac{1}{2} \right ) \right ]-\frac{3}{2}+\frac{1}{t^2} \big \}\nonumber \\
&\times& \{ \frac{\alpha_e}{t} \left [ e^{-2ct|r-x|} - e^{-2ct(r+x)} \right ] +2 r_A e^{2 r_A ct } \nonumber \\
&\times& \left [ E_1 (2ct (|r-x|+r_A)) - E_1 (2ct (r+x+r_A)) \right ] \bigg \} \nonumber \\
\end{eqnarray}
with the orbital quantum number $l$ of the system, $x=(Z-80)\alpha_e$, $r_A= 0.07 Z^2 \alpha_e^3$, and the exponential integral $E_1(r) = 
\int_r^{\infty} ds e^{-s}/s$.

We have determined the nuclear potential and density by assuming a Fermi-charge distribution given by \cite{Hofstadter}
\begin{equation}
\rho_{n}(r)=\frac{\rho_{0}}{1+e^{(r-b)/a}} 
\end{equation}
for the normalization factor $\rho_0$, the half-charge radius $b=5.670729105$ fm \cite{angeli} and $a= 2.3/4(ln3)$ is related to the skin thickness. 

\subsection{RCC theory for unperturbed wave function}

In the RCC theory framework, the unperturbed wave function of an atomic system with a closed-core and a valence orbital like in the case of Cs atom 
due to $H_{em}$ can be expressed as \cite{lindgren,mukherjee}
\begin{eqnarray}
 |\Psi_v^{(0)} \rangle &=& e^{T^{(0)}} \left \{ 1+ S_v^{(0)} \right \} |\Phi_v \rangle ,
\end{eqnarray}
where $|\Phi_v \rangle$ is the reference wave function, which is obtained by solving Dirac-Hartree-Fock (DHF) wave function of the closed-core 
($|\Phi_0 \rangle$) and then, appending the corresponding valence orbital $v$ to it as $|\Phi_v \rangle= a_v^{\dagger}|\Phi_0 \rangle$. $T^{(0)}$ 
and $S_v^{(0)}$ are the core and the valence excitation operators with the superscript 0 represents absence of any external perturbation. The 
amplitudes of the unperturbed RCC operators and energies are obtained by solving the following equations (see, e.g. \cite{bksahoo1,bksahoo2,bksahoo3})
\begin{eqnarray}
  \langle \Phi_0^K | \bar{H}_{em} | \Phi_0 \rangle = \delta_{K,0} E_0^{(0)}  
  \label{eqt0} 
\end{eqnarray} 
and
\begin{eqnarray}
 \langle \Phi_v^M | \bar{H}_{em} \{ 1 + S_v^{(0)} \} | \Phi_v \rangle = E_v^{(0)} \langle \Phi_v^M | \{ \delta_{M,v} + S_v^{(0)} \} | \Phi_v \rangle  , \ \ \ \
 \label{eqamp}
\end{eqnarray}
where $\bar{H}_{em} =e^{-T^{(0)}}H_{em} e^{T^{(0)}}$, the superscripts $K$ and $M$ represent the $K^{th}$ and $M^{th}$ excited state determinants with 
respect to their respective reference states $| \Phi_0 \rangle$ and $| \Phi_v \rangle$, $E_0$ is the energy of the closed-core (i.e. Cs$^+$) and $E_v$ 
is the energy of a neutral state of Cs atom. These energies are determined by
\begin{eqnarray}
 E_0^{(0)} &=& \langle \Phi_0 | \bar{H}_{em} | \Phi_0 \rangle 
 \end{eqnarray}
 and
\begin{eqnarray}
 E_v^{(0)}  &=& \langle \Phi_v | \bar{H}_{em} \left \{ 1+S_v^{(0)} \right \} | \Phi_v \rangle .
 \label{eqeng}
\end{eqnarray} 
$\Delta E_v = E_v^{(0)} - E_0^{(0)}$ is the electron binding energy and is the negative of the electron affinity (EA) 
for the valence $v$ orbital. We have incorporated one-particle and one-hole (single), two-particle and two-hole (double) and three-particle three-hole
(triple) excitations in our calculations through the RCC operators by defining
\begin{eqnarray}
 && T^{(0)} \simeq T_1^{(0)} +  T_2^{(0)} +  T_3^{(0)} \\ 
\text{and} \ \ \ \ \ &&  S_{v}^{(0)} \simeq S_{1v}^{(0)} +  S_{2v}^{(0)} +  S_{3v}^{(0)} ,
\end{eqnarray}
where the subscripts $K$ and $M$ run over 1, 2 and 3 which are referred to as single, double and triple excitations respectively. To assess the 
importance of the triple excitations, we have performed calculations considering single and double excitations in the RCC theory (RCCSD method) 
after exciting all the core electrons, and then with single, double and triple excitations in the RCC theory (RCCSDT method). In addition, we 
have also carried out calculations using the second-order relativistic many-body theory (RMP(2) method), considering two-orders of the residual
interaction and only keeping linear terms from the RCCSD method (RLCCSD method) as 
\begin{eqnarray}
 |\Psi_v^{(0)} \rangle &\simeq& \left \{ 1+ T^{(0)} + S_v^{(0)} \right \} |\Phi_v \rangle .
\end{eqnarray}
Intermediate results from the RMP(2) and RLCCSD methods can demonstrate the propagation of electron correlation effects from lower- to all-order
methods systematically in order to understand the role of electron correlation effects in the accurate calculations of EAs of valence electrons 
in different states of the Cs atom.

\begin{table}[t]
\caption{Calculated EAs (in cm$^{-1}$) at different levels of approximations. Corrections from the Breit and QED interactions are given 
as $\Delta$Breit and $\Delta$QED, respectively. Extrapolated contributions from the finite size basis functions are given as ``Extra'' and 
the estimated uncertainties are quoted within the parentheses.}
\begin{ruledtabular}
\begin{tabular}{lccccc} 
  Method  & $6S$  & $6P_{1/2}$ & $7S$ & $7P_{1/2}$ & $8P_{1/2}$ \\   
 \hline \\ 
 \multicolumn{6}{c}{\underline{Dirac-Coulomb contributions}} \\
DHF       & 27954.01  & 18790.51  & 12111.79  &  9221.90  &  5509.15  \\
RMP(2)  & 31818.40  & 20297.57  & 13026.52  &  9683.05  &  5720.11 \\
RLCCSD    & 31806.94  & 20393.25  & 12936.56  &  9681.38  &  5710.04  \\
RCCSD     & 31520.14  & 20248.86  & 12895.49  &  9647.42  &  5696.17  \\
RCCSDT    & 31347.68  & 20215.57  & 12859.52  &  9639.21  &  5695.68  \\
 \multicolumn{6}{c}{\underline{Corrections from Breit interaction}} \\
 DHF       & $-3.19$  & $-7.49$  & $-1.08$  &  $-2.68$  &  $-1.26$  \\
RMP(2)  & 1.47     & $-6.98$  & $-0.06$  &  $-2.44$  &  $-1.13$ \\
RLCCSD    & 1.08     & $-6.95$  & $-0.38$  &  $-2.45$  &  $-1.13$  \\
RCCSD     & $-0.19$  & $-7.80$  & $-0.54$  &  $-2.57$  &  $-1.20$  \\
RCCSDT & $-0.60$  & $-7.81$  & $-0.65$    & $-2.61$   &  $-1.21$  \\
 \multicolumn{6}{c}{\underline{Corrections from QED interactions}} \\
 DHF       & $-17.25$ & 0.61 & $-4.70$ & 0.22 & 0.10 \\
RMP(2)  & $-24.62$ & 0.68 & $-5.82$ & 0.25 & 0.12 \\
RLCCSD    & $-24.91$ & 0.81 & $-5.62$ & 0.28 & 0.13 \\
RCCSD     & $-22.81$ & 1.25 & $-5.27$ & 0.52 & 0.68 \\
RCCSDT   & $-20.53$ & 1.31    & $-5.09$    &  0.57     &   0.71    \\
\hline \\
Extra     & 30.71    & 14.47 & 6.69 & 3.91  & 2.15 \\
Final     & 31357(50) & 20243(20)  & 12861(15)  & 9641(10)  &  5697(10)  \\
NIST \cite{nist} &  31406.47 & 20229.21 & 12871.94 & 9642.12 & 5698.63 \\
\end{tabular}
\end{ruledtabular}
\label{tab1}
\end{table}

\subsection{RCC theory for perturbed wave function}

Extending the RCC theory {\it ansatz} of atomic wave function, the first-order perturbed wave function due to $H_{w}$ can be expressed 
as \cite{bijaya1,bijaya2,bijaya3}
\begin{eqnarray}
|\Psi_v^{(1)} \rangle &=& e^{T^{(0)}} \left \{ S_v^{(1)}+ T^{(1)} \left (1+S_v^{(0)} \right ) \right \} |\Phi_v \rangle, \label{eqcc} \ \ \ \ \
\end{eqnarray}
where $T^{(1)}$ and $S_v^{(1)}$ are the core and the valence excitation operators with the superscript 1 representing order of perturbation in 
$H_{w}$. After obtaining the amplitudes of the unperturbed RCC operators, we obtain the amplitudes of their perturbed counterparts by
solving the following equations 
\begin{eqnarray}
  \langle \Phi_0^K | \bar{H}_{em} T^{(1)} + \bar{H}_{w} | \Phi_0 \rangle = 0 
  \label{eqt1} 
\end{eqnarray} 
and
\begin{eqnarray}
 \langle \Phi_v^M | (\bar{H}_{em}- E_v^{(0)} ) S_v^{(1)} &+& ( \bar{H}_{em} T^{(1)} + \bar{H}_{w}) \nonumber \\ && \times \{ 1+ S_v^{(0)} \} | \Phi_v \rangle = 0 , \ \ \ \
 \label{eqamp1} 
\end{eqnarray}
where $\bar{H}_{w} = e^{-T^{(0)}} H_{w} e^{T^{(0)}}$. Here, the subscripts $K$ and $M$ run again over 1, 2 and 3 which are referred to as single, 
double and triple excitations respectively. The important difference between the amplitude determining equations for unperturbed and perturbed wave
functions is that the projected determinantal  states (denoted by superscripts $K$ and $M$) have even and odd parities, respectively. The RCC operators
representing perturbed single, double and triple excitations are denoted by
\begin{eqnarray}
 && T^{(1)} \simeq T_1^{(1)} +  T_2^{(1)} +  T_3^{(1)} \\ 
\text{and} \ \ \ \ \ &&  S_{v}^{(1)} \simeq S_{1v}^{(1)} +  S_{2v}^{(0/1)} +  S_{3v}^{(1)} .
\end{eqnarray}
Along with the calculations using the RCCSD and RCCSDT methods, we also determine perturbed wave functions in the RLCCSD approximation
by considering the expression
\begin{eqnarray}
|\Psi_v^{(1)} \rangle &\simeq& \left \{ \left (1+T^{(0)} \right ) S_v^{(1)}+ T^{(1)} \left (1+S_v^{(0)} \right ) \right \} |\Phi_v \rangle.  \ \ \ \ \
\end{eqnarray}
The differences in the results from these methods will demonstrate the role of non-linear in $T^{(0)}$ terms and triple excitations to the amplitudes 
of the first-order perturbed wave functions.

\begin{table}[t]
\caption{Calculated $A_{hyf}$ values (in MHz) from different approximations in the many-body theory are given. Corrections from the Breit interaction, 
QED effect and BW effect are given as $\Delta$Breit, $\Delta$QED and $\Delta$BW, respectively. Estimated ``Extra'' contributions and uncertainties to 
the final calculated values are quoted, but error bars of experimental results are not given because they appear beyond the interested significant 
digits. We have used $g_I=0.737885714$ to determine the theoretical values.}
\begin{ruledtabular}
\begin{tabular}{lccccc} 
  Method  & $6S$  & $6P_{1/2}$ & $7S$ & $7P_{1/2}$ & $8P_{1/2}$ \\   
 \hline \\ 
 \multicolumn{6}{c}{\underline{Dirac-Coulomb contributions}} \\
DHF            & 1433.96  &  161.07  &  394.12  &  57.69  &  27.01  \\
RMP(2)       & 2317.02  &  267.09  &  559.62  &  89.08  &  40.95 \\
RLCCSD         & 2492.22  &  311.80  &  571.67  &  98.45  &  44.21  \\
RCCSD          & 2328.40  &  286.48  &  548.65  &  92.52  &  41.79  \\
RCCSDT         & 2308.52  &  290.21  &  548.48  &  94.03  &  41.65  \\
 \multicolumn{6}{c}{\underline{Corrections from Breit interaction}} \\
DHF       & 0.01  & $-0.68$  & $-0.03$  &  $-0.24$  &  $-0.11$  \\
RMP(2)  & 2.52  & $-0.42$  & 0.54     &  $-0.12$  &  $-0.05$ \\
RLCCSD    & 4.11  & $-0.09$  & 0.75     &  $-0.02$  &  $-0.01$  \\
RCCSD     & 4.71  & $-0.16$  & 0.85     & $-0.03$   &  $-0.02$ \\
RCCSDT  & 4.65     & $-0.18$  &  0.83    & $-0.04$ &  $-0.02$  \\
\multicolumn{6}{c}{\underline{Corrections from QED interactions}} \\
DHF       & $-4.61$ & 0.01  & $-1.18$ & 0.004 & $\sim 0.0$ \\
RMP(2)  & $-7.29$ & 0.05  & $-1.61$ & 0.02  & 0.01 \\
RLCCSD    & $-8.22$ & 0.06  & $-1.69$ & 0.01  & $\sim 0.0$\\
RCCSD     & $-7.58$ & 0.05  & $-1.65$ & 0.01  & $\sim0.0$  \\
RCCSDT    & $-7.28$  & 0.05     & $-1.51$  & 0.01    &  $\sim0.0$ \\
\hline \\
$\Delta$BW & $-6.74$  & $-0.09$  & $-1.62$  & $-0.02$ & $-0.02$ \\
Extra & 7.08 & 0.65  & 0.86 & 0.39 & $\sim 0.0$ \\
Final     & 2306(10) & 291(2)   & 547(2)   & 94(1)   &  42(1)  \\
Experiment & 2298.16$^a$  & 291.91$^b$  & 545.82$^c$ & 94.40$^d$ &  42.97$^e$ \\
\end{tabular}
\end{ruledtabular}
Refs. $^a$\cite{arimondo}; $^b$\cite{dipankar}; $^c$\cite{yang}; $^d$\cite{williams}; $^e$\cite{happer}.\\
\label{tab2}
\end{table}

\begin{table*}[t]
\caption{Magnitudes of the reduced E1 matrix elements in atomic units (a.u.) are given at different levels approximations of many-body theory. Corrections 
from the Breit and QED interactions are given as $\Delta$Breit and $\Delta$QED, respectively, and extrapolated contributions are given as ``Extra''. 
The final values are given along with the uncertainties and compared with the extracted values from the latest experiments.} 
\begin{ruledtabular}
\begin{tabular}{l rrrrrrrrrrr  } 
  Transition  & DHF  & RMP(2) & RLCCSD & RCCSD & RCCSDT  & $\Delta$Breit & $\Delta$QED & Extra & Final & Experiment \\
 \hline \\
$6P_{1/2} \rightarrow 6S$ & 5.2777 & 4.5877 & 4.4740 & 4.5445 &  4.5023 & $-0.0002$ & 0.0011    & 0.0035 & 4.5067(40) & 4.508 \cite{gregoire} \\
$7P_{1/2} \rightarrow 6S$ & 0.3717 & 0.2233 & 0.2962 & 0.2989 &  0.2804 & 0.0006    & $-0.0008$ & 0.0003 & 0.2805(20) & 0.27810 \cite{damitz}\\ 
$8P_{1/2} \rightarrow 6S$ & 0.1321 & 0.8996 & 0.0902 & 0.0919 &  0.0817 & 0.0007    & $-0.0005$ & 0.0005 & 0.0824(10) & \\
$6P_{1/2} \rightarrow 7S$ & 4.4131 & 4.4428 & 4.2025 & 4.2528 &  4.2510 & 0.0041    & $-0.0017$ & 0.0025 & 4.2559(30) & 4.249 \cite{Toh} \\ 
$7P_{1/2} \rightarrow 7S$ & 11.0121& 10.2646& 10.2481& 10.2921& 10.2795 & $-0.0015$ & 0.0025 & 0.0110 & 10.2915(100)  & 10.308 \cite{Bennett1} \\
$8P_{1/2} \rightarrow 7S$ & 0.9336 & 0.9437 & 0.9431 & 0.9501 &  0.9602 & 0.0028    & $-0.0015$ & 0.0008 & 0.9623(20) & \\
 \end{tabular}
\end{ruledtabular}
\label{tab3}  
\end{table*}

\subsection{Evaluation of atomic properties}

To test the accuracies of the wave functions, we also evaluate other relevant properties apart from the binding energies and compare them with their 
high precision experimental values. The accuracies of the calculated energies are sensitive to the quality of the wave functions slightly away from 
the nuclear region of atomic systems. For testing the accuracies of the wave functions in the nuclear region and the far nuclear region, we 
evaluate the magnetic dipole hyperfine structure constants ($A_{hf}$) and the electric dipole (E1) transition amplitudes, and compare them with their 
respective experimental values. These quantities were evaluated using the expression
\begin{eqnarray}
 \langle O \rangle_{fi} &=& \frac{ \langle \Psi_f^{(0)} | O | \Psi_i^{(0)} \rangle} {\sqrt{\langle \Psi_f^{(0)} | \Psi_f^{(0)} \rangle \langle \Psi_i^{(0)} | \Psi_i^{(0)} \rangle}} \nonumber \\
           &=& \frac{\langle \Phi_f | \{S_f^{(0)\dagger} +1 \} \overline{O} \{ 1+ S_i^{(0)} \} |\Phi_i \rangle} {\langle \Phi_f | \{S_f^{(0)\dagger} +1 \} \overline{N} \{ 1+ S_i^{(0)} \} |\Phi_i \rangle},
\label{expv}
\end{eqnarray}
where $\overline{O}= e^{T^{(0)\dagger}} O e^{T^{(0)}}$ for the operator $O$ representing the respective property and $\overline{N}= e^{T^{(0)\dagger}} 
e^{T^{(0)}}$. In the evaluation of $A_{hyf}$, we set $| \Psi_f^{(0)} \rangle = |\Psi_i^{(0)} \rangle$. Both $\overline{O}$ and $\overline{N}$ are the 
non-terminating series, which are evaluated by adopting iterative procedures as described in Refs. \cite{bksahoo1,bksahoo2,nandy}. We also present
results from the RMP(2) and RLCCSD methods to make a comparative analysis of trend of correlation effects in the determination of the 
aforementioned properties. We have used $g_I=\mu_I/I=0.737885714$ with nuclear magnetic moment $\mu_I$ for the evaluation of $A_{hyf}$
values. We have also taken into account the Bohr-Weisskopf (BW) effect by defining the nuclear magnetization function ($F(r)$) in the Fermi nuclear charge 
distribution approximation as  
\begin{eqnarray}
 F(r) &=& \frac{f_{WS}}{\cal N} [ (r/b)^3 -3(a/b)(r/b)^2 R_1((b-r)/a) \nonumber \\ && + 6 (a/b)^2(r/b)R_2((b-r)/a) - 6 (a/b)^3 \nonumber \\ 
 && \times R_3((b-r)/a) +6(a/b)^3R_3(b/a) ]
 \label{magd}
\end{eqnarray}
for $r \le b$ and
\begin{eqnarray}
 F(r) &=& 1- \frac{1}{\cal N} [ 3 (a/b) (r/b)^2 R_1((r-b)/a) \nonumber \\ && + 6 (a/b)^2(r/b)R_2((r-b)/a)]
\end{eqnarray}
for $r>b$, where 
\begin{eqnarray}
 {\cal N} = 1 + (a/b)^2 \pi^2 + 6 (a/b)^3 R_3(b/a)
\end{eqnarray}
and 
\begin{eqnarray}
 R_k(x) = \sum_{n=1}^{\infty} (-1)^{n-1} \frac{e^{-nx}}{n^k} . 
\end{eqnarray}
In Eq. (\ref{magd}), $f_{WF}$ takes into account the Woods-Saxon (WS) potential correction and is estimated after neglecting the spin-orbit 
interaction within the nucleus using the following expressions \cite{shabaev1,shabaev2}
\begin{eqnarray}
 f_{WS} &=& 1- \left (\frac{3}{\mu_I} \right ) ln \left ( \frac{r}{b} \right ) \left [- \frac{2I-1}{8(I+1)}g_S + (I-1/2)g_L \right ]   \nonumber
\end{eqnarray}
for $I=L+\frac{1}{2}$ and
\begin{eqnarray}
 f_{WS} &=& 1- \left (\frac{3}{\mu_I} \right ) ln \left ( \frac{r}{b} \right ) \left [ \frac{2I+3}{8(I+1)}g_S + \frac{I(2I+3)}{2(I+1)}g_L \right ]   \nonumber
\end{eqnarray}
for $I=L-\frac{1}{2}$ with the total orbital angular momentum $L$ of the nucleus. We have used the nuclear parameters $g_L=1$ and $g_S=4.143$ for $^{133}$Cs atom \cite{shabaev1}. 

\subsection{Evaluation of $E1_{PV}$}

In the RCC theory framework, Eq. (\ref{e1pnc}) is given by
\begin{eqnarray}
E1_{PV} &\simeq& \frac{\langle \Phi_f | \{S_f^{(1)} + (S_f^{(0)\dagger} +1) T^{(1)\dagger}\} \overline{D} \{ 1+ S_i^{(0)} \} |\Phi_i \rangle} {\langle \Phi_f | \{S_f^{(0)\dagger} +1 \} \overline{N} \{ 1+ S_i^{(0)} \} |\Phi_i \rangle} \nonumber \\
 + && \frac{\langle \Phi_f |\{ S_f^{(0)\dagger} +1 \} \overline{D} \{T^{(1)}(1+ S_i^{(0)}) + S_i^{(1)}\} |\Phi_i \rangle}{\langle \Phi_f | \{S_f^{(0)\dagger} +1 \} \overline{N} \{ 1+ S_i^{(0)} \} |\Phi_i \rangle} , \ \ \
\end{eqnarray}
where $\overline{D}= e^{T^{(0)\dagger}} D e^{T^{(0)}}$ and $\overline{N}= e^{T^{(0)\dagger}} e^{T^{(0)}}$. Contributions from the non-terminating 
expressions $\overline{D}$ and $\overline{N}$ are estimated by an iterative approach similar to that used in the expression for evaluating
 properties, which is given in 
Eq. (\ref{expv}). The ``Core'' contributions for the initial and final states originate from $T^{(1)\dagger} \overline{D}$ and $\overline{D}T^{(1)}$ 
respectively, and the rest of the RCC terms involving $S_f^{(0/1)\dagger}$ and $S_i^{(0/1)}$ give rise to valence contributions from the `final' and 
`initial' states, respectively. The simultaneous presence of both the electromagnetic and NSI weak interactions through the RCC operators account for core
and valence correlation contributions, including the DCP correlation effects.  

\subsection{Basis functions}

 We have used Gaussian type orbitals (GTOs) \cite{Boys} to construct the single particle DHF wave functions. The radial components for the large and 
small components of DHF orbitals are expressed using these GTOs as
\begin{eqnarray}
 P(r) = \sum_{k=1}^{N_k} c_k^{\cal L} \zeta_{\cal L} r^l e^{-\alpha_0 \beta^k r^2}
\end{eqnarray}
and 
\begin{eqnarray}
 Q(r) = \sum_{k=1}^{N_k} c_k^{\cal S} \zeta_{\cal L} \zeta_{\cal S} \left (\frac{d}{dr} + \frac{\kappa}{r}\right ) r^l e^{-\alpha_0 \beta^k r^2} ,
\end{eqnarray}
where $P(r)$ and $Q(r)$ are the large and small radial components of the DHF orbitals, $l$ is the orbital quantum number, $\kappa$ is the 
relativistic angular momentum quantum number, $c_k^{{\cal L}({\cal S})}$ are the expansion coefficients, $\zeta_{{\cal L}({\cal S})}$ are the normalization factors of
GTOs, $\alpha_0$ and $\beta$ are optimized GTO parameters for a given orbital, and $N_k$ represents the number of GTOs used. We have considered
40 GTOs for each symmetry up to $l=6$ for the RCC calculations and up to $l=9$ for analyzing results using the RMP(2) method. For the construction 
of GTOs, the values of $\alpha_0$ we use are 0.0009, 0.0008, 0.001, 0.004 and 0.005 for the $s$, $p$, $d$, $f$ and other higher angular momentum symmetry orbitals, respectively. The
corresponding $\beta$ values we have used are  2.15, 2.15, 2.15, 2.25 and 2.35 for the $s$, $p$, $d$, $f$ and other higher symmetry orbitals, respectively.
Since our orbitals are not bounded by a cavity, we carry out the numerical integration of radial integrals up to $r=500$ a.u. using a 10-point 
Newton-Cotes Gaussian quadrature formula on grids. Non-linear grids are defined, as in Ref. \cite{grasp2}, for the numerical calculations with the 
step-size 0.0199 a.u. over 1200 grid points. We have considered excitations from all the occupied orbitals,  but limited the virtual space to excitations 
of orbitals in that space with energies less than 2000 a.u. This includes 1$-$19$s$, 2$-$19$p$, 3$-$19$d$, 4$-$18$f$, 5$-$16$g$, 6$-$15$h$ and 7$-$15$i$-symmetry orbitals. These orbitals will be
referred to as the ``active orbitals'' hereafter. 

\begin{table}[t]
\caption{List of the experimental values of E1 matrix elements (in a.u.) for a few low-lying transitions reported over the years using different 
measurement techniques.}
\begin{ruledtabular}
\begin{tabular}{l ccc} 
 Transition &  Value & Reference & Year \\
 \hline \\
$6P_{1/2} \leftrightarrow 6S$ & 4.5097(74) & \cite{Young}   & 1994 \\
                              & 4.4890(65) & \cite{Rafac}   & 1999 \\
                              & 4.505(2)   & \cite{Patternson} & 2015 \\
                              & 4.508(4)   & \cite{gregoire} & 2015 \\
$7P_{1/2} \leftrightarrow 6S$ & 0.2825(20) & \cite{Vasilyev} & 2002 \\
                              & 0.2789(16) & \cite{Antypas} &  2013 \\
                              & 0.27810(45) & \cite{damitz} & 2019 \\
$8P_{1/2} \leftrightarrow 6S$ &  Not available yet  &                           \\
$6P_{1/2} \leftrightarrow 7S$ & 4.233(22) & \cite{Bouch} & 1984 \\
                              & 4.249(4) & \cite{Toh} & 2019 \\
$7P_{1/2} \leftrightarrow 7S$ & 10.308(15) & \cite{Bennett1} & 1999 \\
$8P_{1/2} \leftrightarrow 7S$ & Not available yet &   \\
 \end{tabular}
\end{ruledtabular}
\label{tab31}
\end{table}

\section{Results and Discussion}

\subsection{Cs APV calculations \& context of present work}
 
  The main thrust of our present work is a high-precision calculation of $E1_{PV}$ for the $6s ~ ^2S_{1/2} - 7s ~ ^2S_{1/2}$ in $^{133}$Cs; the transition on which 
the most accurate APV measurement (0.35\% accuracy) has been carried out to date \cite{wood}. As mentioned earlier, a considerable amount of effort has 
also gone into performing very accurate calculations on $E1_{PV}$ using state-of-the art relativistic many-body theories (e.g. see 
\cite{porsev1,dzuba1} and references {\it therein}). At the time of the last Cs APV measurement, the accuracies of the atomic calculations were about 
1\% \cite{dzuba2,blundell}. Later by using the amended values of the E1 matrix elements inferred from the high precision measurements of 
lifetimes and polarizabilities of atomic states, the uncertainty in the calculation was reduced to 0.4\% \cite{bennett}. This yielded a $Q_W^{\rm at}$ 
that disagreed with its SM value by 2.5$\sigma$. Subsequently, the leading order relativistic correction from the Breit interaction and the lower-order
QED effects and the neutron skin were included in the atomic calculations (refer to \cite{kozlov,shabaev,Derevianko3} for discussions). As pointed 
out before, there has been a renewed interest in the inclusion of the neglected correlation effects in Cs APV since about a decade.
(e.g. see discussions in \cite{derevianko1,derevianko2}). The latest calculations including the effect of the valence triple excitations were 
investigated by employing the RCC theory, and it was found that their contributions to the atomic properties of $^{133}$Cs were relatively important 
in reducing the uncertainty in the $E1_{PV}$ amplitude to 0.27\% \cite{porsev1}. This result is in good agreement with the SM, however the
calculation on which it is based had used a sum-over-states approach in which the leading contributions from the excited states up to the principal quantum number $n=9$ 
were estimated by using matrix elements, calculated using the RCC theory and referred to as ``Main'' contribution. The rest were classified 
into ``Core'' and ``Tail'', and they were evaluated using mixed many-body methods \cite{porsev1}. Later, Dzuba {\it et al.} reported  another result with 0.5\% accuracy by evaluating the 
``Main'' contribution, again, using a sum-over-states approach but with different ``Core'' (opposite sign than \cite{porsev1}) and ``Tail'' 
contributions by taking into account certain sub-classes of correlation effects \cite{dzuba1}. This resulted in a difference of about 0.8\% 
between the $E1_{PV}$ calculations of Porsev {\it et al} \cite{porsev1} and Dzuba {\it et al}. \cite{dzuba1}. Following these works, Roberts 
{\it et al.} have reported  the contributions from QED and DCP effects \cite{roberts2,Roberts}. There are still unresolved issues in the determination of 
electron correlation in Cs APV due to the disparate approaches that have been used in the treatment of different physical effects in the low- and 
high-lying excited states. In other words, the ``Main", ``Tail" and ``Core" contributions have not been evaluated on par with each other. Also, the 
Breit interaction and the effective QED interactions have not been treated at the same level as the DC interaction in Refs. \cite{dzuba1,porsev1}. 
The contributions from the triple excitations involving core orbitals were not determined in Ref. \cite{porsev1}. In contrast to the previous
previous works, our calculation of the $E1_{PV}$ amplitude adopts an approach based on the perturbed RCC theory as outlined above. We excite all the core electrons in our RMP(2), RLCCSD and RCCSD calculations to account for the electron correlation effects. However, we correlate all the 
electrons except the $1-3s$, $2-3p$, and $3d$ occupied orbitals and beyond $n=15$ virtual orbitals for triple excitations due to limitations in
the available computational resources.

\begin{table*}[t]
\caption{Magnitudes of the $H_{APV}^{NSI}$ matrix elements in $-i (Q_{W}/N) \times 10^{-11}$ from different methods. Corrections from the Breit and 
QED interactions are given as $\Delta$Breit and $\Delta$QED, respectively. The final values after including ``Extra'' contributions are given 
along with the uncertainties in the parentheses. The quantity ${\cal X}$ is defined as ${\cal X} = |[{\cal R}_{th}/{\cal R}_{ex}]-1| \times \langle
\Psi_k^{(0)} || H_{APV}^{NSI}|| \Psi_v^{(0)} \rangle$ for the corresponding theoretical (denoted with subscript $th$) and experimental (denoted with 
subscript $ex$) values, where ${\cal R} = \sqrt{A_{hyf}^{k} A_{hyf}^v}$ with superscripts $k$ and $v$ designated for the states with valence orbitals
$k$ and $v$, respectively.} 
\begin{ruledtabular}
\begin{tabular}{l rrrrrrrrrc  } 
  $| \Psi_k^{(0)} \rangle \rightarrow | \Psi_v^{(0)} \rangle$  & DHF  & RMP(2) & RLCCSD & RCCSD  & RCCSDT  & $\Delta$Breit & $\Delta$QED & Extra & Final & ${\cal X}$ \\
 \hline \\
$6P_{1/2} \rightarrow 6S$ & 0.7286 & 1.1955 & 1.3536 & 1.2567 & 1.2725 & $-0.0066$ & $-0.0055$ & 0.0010 & 1.2648(15) & 0.0002 \\
$7P_{1/2} \rightarrow 6S$ & 0.4362 & 0.6909 & 0.7628 & 0.7164 & 0.7268 & $-0.0036$ & $-0.0030$ & 0.0008 & 0.7210(15) & 0.0004 \\
$8P_{1/2} \rightarrow 6S$ & 0.2985 & 0.3782 & 0.5117 & 0.4823 & 0.4821 & $-0.0024$ & $-0.0021$ & 0.0007 & 0.4783(10) & 0.0010 \\
$6P_{1/2} \rightarrow 7S$ & 0.3820 & 0.5891 & 0.6465 & 0.6094 & 0.6205 & $-0.0032$ & $-0.0022$ & 0.0010 & 0.6161(15) & 0.0005 \\
$7P_{1/2} \rightarrow 7S$ & 0.2287 & 0.3393 & 0.3624 & 0.3458 & 0.3493 & $-0.0019$ & $-0.0015$ & 0.0005 & 0.3464(10) & 0.0010 \\
$8P_{1/2} \rightarrow 7S$ & 0.1565 & 0.2117 & 0.2425 & 0.2321 & 0.2314 & $-0.0012$ & $-0.0009$ & 0.0003 & 0.2296(05) & 0.0103 \\
 \end{tabular}
\end{ruledtabular}
\label{tab4}
\end{table*}

\subsection{Ancillary Properties}

 At the outset, we would like to reemphasize that it is customary to compare the results of the calculations of energies, E1 matrix 
elements and $A_{hyf}$ values based on a many-body theory with the available experimental data to assess the accuracy of the $E1_{PNC}$ amplitude. 
We give values for all these quantities by taking into account contributions from the DC Hamiltonian, the Breit interaction, and the QED effects
at different levels of approximation in the many-body methods systematically. We have also estimated the contributions to different properties by
extrapolating our basis functions to infinite-size, which we have referred to as ``Extra", and  given their values. The uncertainties in our 
calculations are estimated by analyzing the optimized GTOs used in the calculations and contributions from the higher level excitations that
are neglected here.

\begin{table}[t]
\caption{Correlation contribution to the $E1_{PV}$ amplitude (in $-i (Q_{W}/N) ea_0 \times 10^{-11}$) of the $6s ~ ^2S_{1/2} - 7s ~ ^2S_{1/2}$ 
transition in $^{133}$Cs from different terms of the RLCCSD, RCCSD and RCCSDT methods. `Others' are the terms including correction due to 
normalization of wave functions that are not mentioned explicitly. Contributions corresponding to ``Core'' and ``Valence'' correlations are 
given separately to distinguish them. $\overline{D} \equiv D$ in the RLCCSD method approximation.}
\begin{ruledtabular}
\begin{tabular}{lrrr}
 RCC term & RLCCSD & RCCSD & RCCSDT  \\
\hline
\multicolumn{4}{c}{Core contributions} \\
 $\overline{D}T_1^{(1)}$  & $-0.0534$  & $-0.0410$ & $-0.0410$ \\
 $T_1^{(1)\dagger} \overline{D}$  & 0.0519  & 0.0392  & 0.0392   \\
 Others  & $-0.0001$ &  $-0.0001$  & $\sim 0.0$ \\    
 \hline 
 Total                                            & $-0.0016$ & $-0.0019$ & $-0.0018$ \\
 \hline \hline \\
\multicolumn{4}{c}{Valence (Main$+$Tail) contributions} \\
 $\overline{D}S_{1i}^{(1)}$                       & $-0.1663$ & $-0.1913$ & $-0.1874$  \\
 $S_{1f}^{(1)\dagger} \overline{D}$               & 2.0603    & 1.8064    &  1.7925    \\
 $S_{1f}^{(0)\dagger} \overline{D} S_{1i}^{(1)}$  & $-0.3045$ & $-0.2336$ & $-0.2288$   \\
 $S_{1f}^{(1)\dagger} \overline{D} S_{1i}^{(0)}$  & $-0.5529$ & $-0.4218$ & $-0.4147$   \\
 $\overline{D}S_{2i}^{(1)}$                       & $-0.0357$ & $-0.0263$ & $-0.0257$  \\
 $S_{2f}^{(1)\dagger} \overline{D}$               & 0.0006    & 0.0009    & 0.0004     \\
 $T_{2}^{(0)\dagger} D S_{3i}^{(1)}$              &           &           & $-0.0019$  \\
 $S_{3f}^{(1)\dagger} D T_{2}^{(0)}$              &           &           & $-0.0007$  \\
 $T_{2}^{(1)\dagger} \overline{D} S_{3i}^{(0)} $  &           &           & $-0.0004$ \\
 $S_{3f}^{(0)\dagger} \overline{D} T_{2}^{(1)}$   &           &           & $-0.0006$  \\
 $S_{2f}^{(0)\dagger} \overline{D} S_{3i}^{(1)}$  &           &           & $-0.0006$  \\
 $S_{3f}^{(1)\dagger} \overline{D} S_{2i}^{(0)}$  &           &           & 0.0007    \\
 Others                                           & $-0.0608$ & $-0.0363$ & $-0.0343$  \\
 \hline 
 Total                                            & 0.9407 & 0.8980   & 0.8985     \\ 
  \end{tabular}
\end{ruledtabular}
\label{tab5}
\end{table}

In Table \ref{tab1}, we give the final EA values from our calculations and these values are compared with the 
precise measurements listed in the National Institute of Science and Technology (NIST) database \cite{nist}. Following this, we have given the 
$A_{hyf}$ values in Table \ref{tab2} using different methods. After adding up all the contributions along with corrections from the BW effect, the 
final values are compared with the high-precision experimental values \cite{arimondo,dipankar,yang,williams,happer}. It can be seen that the triple 
excitations improve the $A_{hyf}$ results of the $P_{1/2}$ states quite significantly. Contributions from the Breit and QED interactions are 
non-negligible for achieving high precision results. We give the values of the reduced matrix elements of $D$ of important transitions along with 
their error bars in Table \ref{tab3}. The extracted E1 matrix elements from the latest precise measurements of lifetimes and Stark shifts are given in the 
same table. The agreement between our calculations and the experimental values \cite{Young,Vasilyev,Toh,Bennett1} is found to be quite good. However,
we would like to mention that the experimental values of these matrix elements have been reported differently over the time \cite{Young,Vasilyev,
Toh,Bennett1,Rafac,Patternson,gregoire,Antypas,damitz,Bouch}; sometime they do not even agree within the quoted error bars as  can be found from
the list given in Table \ref{tab31}. Nonetheless, it can be seen from Table \ref{tab3} that the DHF values of E1 matrix elements are large in magnitude and they reduce 
successively after the inclusion of the correlation effects at the RCCSD and RCCSDT levels. The triples contributions to the E1 matrix elements are 
more significant than those in the case of other properties for $^{133}$Cs. Similarly, the matrix elements of $H_{APV}^{NSI}$ are given in Table 
\ref{tab4}. As can be seen from this table, the correlation trends in the matrix elements of $H_{APV}^{NSI}$ are completely different than those for 
the E1 matrix elements but almost similar to those of $A_{hyf}$. We analyze the accuracies of ${\cal R} = \sqrt{A_{hyf}^{k} A_{hyf}^v}$, the 
superscripts $k$ and $v$ denoting for states with valence orbitals $k$ and $v$ respectively, by comparing our theoretical values with the experimental
results. This is used to determine the accuracy of the $\langle \Psi_k | H_{APV}^{NSI}|\Psi_v\rangle$ matrix elements and their accuracies are 
quantified by evaluating ${\cal X} = |[{\cal R}_{th}/{\cal R}_{ex}]-1| \times \langle \Psi_k^{(0)} || H_{APV}^{NSI}|| \Psi_v^{(0)} \rangle$ values, 
with subscripts $th$ and $ex$ referring to our theoretical values and experimental results respectively, for important low-lying states. These values 
are found to be very small, implying that the $H_{APV}^{NSI}$ matrix elements are obtained quite accurately by us.

\begin{table}[t]
\caption{Contributions from the `Core' and `Valence' correlations to the $E1_{PV}$ amplitude (in $-i (Q_{W}/N) ea_0 \times 10^{-11}$)
using the Dirac-Coulomb Hamiltonian in the DHF, RCCSD and RCCSDT methods. Valence contributions are given in two parts as `Main' by considering 
contributions only from the $np ~ ^2P_{1/2}$ states with $n=6$, 7 and 8, while `Tail' refer to the contributions from the remaining bound states 
and continuum. Contributions from the extrapolated basis function, ``Extra'' and neutral weak interactions among electrons ($e-e$) are also quoted.}
 \begin{ruledtabular}
  \begin{tabular}{lccccc }
  Method    & Core & Main & Tail &  Extra & $e-e$ \cite{Milstein} \\
 \hline \\
 DHF & $-0.0017$  & 0.7264 & 0.0137  &  & \\ 
 RCCSD & $-0.0019$ & 0.8623 & 0.0357 &  & \\
 RCCSDT & $-0.0018$ & 0.8594 & 0.0391 & 0.0026 & 0.0003  \\
 \hline \\ 
 Ref. \cite{blundell}$^{\dagger}$ & $-0.002(2)$ & 0.893(7)   &  0.018(5) & \\
 Ref. \cite{porsev1}$^{\dagger}$ & $-0.0020$ & 0.8823(17) & 0.0195 & \\
 Ref. \cite{dzuba1}$^{\dagger}$ & $0.0018(8)$ & 0.8678 & 0.0238(35) & \\
  \end{tabular}
 \end{ruledtabular}
 \label{tab51}
 $^{\dagger}$ Contains additional contribution from the $9p ~ ^2P_{1/2}$ state.
\end{table}

\subsection{$E1_{PV}$ results}

In Table \ref{tab5}, we present and compare our $E1_{PV}$ results for the $6s ~ ^2S_{1/2} - 7s ~ ^2S_{1/2}$ transition in $^{133}$Cs from 
different terms of the RLCCSD, RCCSD and RCCSDT approximations. For the sake of brevity, we present contributions from terms representing ``Core'' 
correlations and valence correlations separately in the same table. It should be noted that these valence correlation contributing terms contain 
both ``Main'' and ``Tail'' contributions of the sum-over-states approach implicitly. As can be seen from the table, the RLCCSD result seems to 
be relatively large, but the rather small difference between the RCCSD and RCCSDT values suggests the convergence of the results after the 
inclusion of higher level particle-hole excitations. The fairly large RLCCSD value is not entirely surprising, given that in this method there have 
been quite significant deviations of various 
spectroscopic properties from their experimental values as discussed in the previous subsection. The differences in the spectroscopic properties at the RCCSD and RCCSDT levels are somewhat large,  and their trends are nonuniform. For example, it can be seen from Tables \ref{tab1} and \ref{tab3} that the calculated energies and E1 matrix elements
decrease in going from the RCCSD method to the RCCSDT method, while the matrix elements of $H_{APV}^{NSI}$, given in Table \ref{tab4}, increase. This is the reason for the small difference between the RCCSD and RCCSDT  $E1_{PV}$ values.
It can be seen from Table \ref{tab5} that there are significant changes in the core contributions through the individual RCC terms in the 
RLCCSD and RCCSD methods, but the differences in the RCCSD and RCCSDT methods are negligibly small. However, we find that changes in the valence 
correlations from different RCC terms in all the three approximations are relatively large. Compared to contributions from the first-order 
perturbed $\overline{D}S_{1i}^{(1)}$ term of the ground state, the perturbed $S_{1f}^{(1)\dagger}\overline{D}$ term of the excited $7S_{1/2}$ 
state contributes predominantly, which correspond to contributions mainly from the one-particle one-hole excitations. The contributions from 
the two-particle two-hole excitations to $E1_{PV}$ are found to be small, which are represented by $\overline{D}S_{2i}^{(1)}$ and $S_{2f}^{(1)\dagger} \overline{D}$ 
for the perturbed wave functions of the ground and excited states, respectively. As mentioned above, there is a small difference between the final 
results from the RCCSD and RCCSDT methods. However, a comparison of the contributions of the individual terms obtained from both these methods reveals that there are significant differences among them. This is because the RCCSD wave functions change when triple excitations are added, due to
the change in the coupled cluster amplitudes. However, this change leads to large cancellations among the net contributions of the individual
terms arising through the initial and final perturbed wave functions resulting in a small difference in their final values. This is also in accordance 
with our analysis of energies and E1 matrix elements changing differently than the matrix elements of $H_{APV}^{NSI}$ in both the methods, which
are manifested in the contributions from the individual RCC terms in a different form. Nonetheless, the convergence of $E1_{PV}$ amplitude with the 
higher-level excitations in the framework of the  RCC theory strongly suggests that the neglected correlation effects are indeed small.

By using the calculated energies, E1 matrix elements and amplitudes of $H_{APV}^{NSI}$ for the intermediate $n(=6,7,8)P_{1/2}$ states at 
different levels of approximations in the tables previously discussed, we estimated the ``Main'' contributions for a qualitative comparison 
of its value with other results reported using the sum-over-states approach. Combining the ``Main'' contributions with the ``Core'' contributions, 
contributions from the ``Tail'' are estimated in the DHF, RCCSD and RCCSDT methods. This breakdown from the DHF, RCCSD and RCCSDT methods are given in 
Table \ref{tab51} and compared with the previously reported values from the sum-over-states approach. Our core contributions are in agreement 
with the values reported in \cite{porsev1,blundell}, but it differs from the latest calculation reported in \cite{dzuba1}. Since the contribution from the 
the $9P_{1/2}$ state is not included in our ``Main'' contribution and contained in the``Tail'', it would be more appropriate to make comparison among the 
total valence correlation contributions (``Main$+$Tail'') from different calculations. We find that our valence correlation contributions are 
0.8980 and 0.8985 from the RCCSD and RCCSDT methods, respectively, against the values 0.911 \cite{blundell}, 0.9018 \cite{porsev1}, and 0.8916 
\cite{dzuba1} in units of $ \times 10^{-11} i (-Q_{W}/N) ea_0$. This shows that our valence correlation contribution is closer to that of 
\cite{dzuba1}. In Table \ref{tab51}, we also present contributions from the extrapolated basis functions, denoted as
``Extra", and a small contribution to $E1_{PV}$ from Ref. \cite{Milstein} due to possible neutral weak interactions among electrons ($e-e$) that
was not included in our calculation.  

\begin{table}[t]
\caption{Comparison of contributions from the Breit and QED interactions to the $E1_{PV}$ amplitude (in $-i (Q_{W}/N) ea_0 \times 10^{-11}$) of 
the $6s ~ ^2S_{1/2} - 7s ~ ^2S_{1/2}$ transition in $^{133}$Cs from various methods employed by different works.}
 \begin{ruledtabular}
  \begin{tabular}{cccc}
 Breit & QED &  Method  &  Reference \\
 \hline \\
$-0.0055(5)$ & $-0.0028(3)$  & RCCSDT   & This work \\
      & $-0.0029(3)$  & Correlation potential &  Ref. \cite{Flambaum}   \\
  $-0.0054$ &  & RMP(3)  &  Ref. \cite{Derevianko3} \\
 $-0.004$ &  & Optimal energy &   Ref. \cite{kozlov} \\
  &   $-0.33(4)$\%   & Radiative potential & Ref. \cite{Roberts} \\
   $-0.0055$  &  & Correlation potential  & Ref. \cite{Dzubab}\\
  $-0.0045$ & $-0.27(3)$\%  & Local DHF potential & Ref. \cite{shabaev} \\
  \end{tabular}
 \end{ruledtabular}
 \label{tab52}
\end{table}

In Table \ref{tab52}, we also give contributions from the Breit and QED interactions using the RCCSDT method and compare them with the values 
reported by other approaches earlier \cite{Flambaum,Derevianko3,kozlov,Roberts,Dzubab,shabaev}. We have also mentioned the many-body method 
employed by other works in the same table to estimate contributions from the Breit and QED interactions to $E1_{PV}$ of the 
$6s ~ ^2S_{1/2} - 7s ~ ^2S_{1/2}$ transition in $^{133}$Cs. We find consistency in the results obtained from various works. This means that 
these relativistic corrections are not influenced significantly by the electron correlation effects. Nonetheless, our method is more 
rigorous than the previous calculations of these corrections to the above $E1_{PV}$ amplitude.

After taking into account contributions from the DC Hamiltonian, Breit interaction and QED effects from the RCCSDT method, the estimated value of 
``Extra'' and small correction from the $e-e$ contribution, we obtain the $E1_{PV}$ amplitude of the $6s ~ ^2S_{1/2} - 7s ~ ^2S_{1/2}$ transition 
in $^{133}$Cs as $0.8914 \times 10^{-11} i (-Q_W/N) ea_0$. To estimate its uncertainty, we adopt the following approach: We have taken the 
difference between the RCCSD and RCCSDT values to estimate the uncertainties in the core and valence contributions to $E1_{PV}$. The major source 
of error for this transition amplitude comes from the finite size of the basis used in our calculation, which is extrapolated to be 
$0.0026 \times 10^{-11} i (-Q_W/N) ea_0$. We assume this as the maximum uncertainty arising from the incomplete basis functions. This approach to the
estimation of the error is more rigorous than the one adopted in Ref. \cite{porsev1}. In the latter work, an uncertainty of 
10\% is assigned to the ``Core'' and ``Tail'' contributions based on the spread of their results in different approximations, and the uncertainty in 
``Main'' is taken to be 0.18\% by analyzing results from a calculation using an {\it ab initio} calculation and another obtained from scaled wave 
functions. We have also estimated the uncertainties from the Breit and QED contributions. Adding all the uncertainties mentioned above in quadrature,
we find that the final uncertainty in $E1_{PV}$ is $0.0027\times 10^{-11} i (-Q_W/N) ea_0$. 

\begin{table}[t]
\caption{Progresses in the atomic calculation of the $E1_{PV}$ amplitude (in $-i (Q_{W}/N) ea_0 \times 10^{-11}$) of the 
$6s ~ ^2S_{1/2} - 7s ~ ^2S_{1/2}$ transition in $^{133}$Cs  over the years by adopting various approaches.}
\begin{ruledtabular}
\begin{tabular}{llcc}
 Year & Result & Approach & Reference  \\
\hline
1989  & 0.908(9) & {\it Ab initio} & Ref. \cite{dzuba2} \\
1990  & 0.909(4) & Sum-over-states & Ref. \cite{blundell}  \\
2000  & 0.8991(36) & Ref. \cite{blundell} + Breit & Ref. \cite{Derevianko3} \\
2001  & 0.901 & Scaled optimal energy & Ref. \cite{kozlov}\\
2002  & 0.904(5) & {\it Ab initio} & Ref. \cite{dzuba0} \\
2005  & 0.904 &  Ref. \cite{dzuba0}$+$QED corr. & Ref. \cite{shabaev} \\
2009  & 0.8906(24) & Sum-over-states & Ref. \cite{porsev1} \\
2012 & 0.8977(40) & Ref. \cite{porsev1}$+$core corr. & Ref. \cite{dzuba1} \\
2020  & 0.8914(27) & {\it Ab initio} & This work \\
  \end{tabular}
\end{ruledtabular}
\label{tab6}
\end{table}

 We have given a list of the calculated $E1_{PV}$ amplitude of the $6s ~ ^2S_{1/2} - 7s ~ ^2S_{1/2}$ transition in $^{133}$Cs over the years in 
Table \ref{tab6}. We also mention the approaches used in the previous works to determine this quantity. As can be seen, apart from a few 
calculations, most of the previous results were reported either using the sum-over-states approach or by considering mixed many-body methods.
The last two high-precision calculations were carried out by adopting the sum-over-states approach, and estimating ``Core'' and ``Tail'' contributions 
using different types of many-body methods. Our {\it ab initio} calculation has similar accuracy to those are obtained using 
the sum-over-states approach, but our error estimation is more rigorous than that of the latter. The most important feature of our work is that it 
treats correlation contributions from the ``Core'', ``Main'' and ``Tail'' sectors at par with each other, thereby resolving the large discrepancy in 
the ``Core'' contribution between the works reported in Refs. \cite{dzuba1} and \cite{porsev1} in an unambiguous manner.

\begin{table}[t]
\caption{Contributions to the scalar dipole polarizability ($\alpha$) of the $6s ~ ^2S_{1/2} - 7s ~ ^2S_{1/2}$ transition in $^{133}$Cs 
using the most precise E1 matrix element amplitudes from the available measurements and our calculations. We have used experimental energies from the 
NIST database \cite{nist} to reduce the uncertainty in the result. Estimated uncertainties from the E1 matrix matrix elements are quoted within the
parentheses.}
 \begin{ruledtabular}
  \begin{tabular}{lccc}
 Intermediate & Initial state  & Final state  & Contribution \\
 state & $6s ~ ^2S_{1/2}$   & $7s ~ ^2S_{1/2} $ & (in a.u.) \\
 \hline \\
$\rightarrow 6p ~ ^2P_{1/2}$  & 4.5067(40) & $-4.2559(30)$ & $-32.60(6)$ \\
$\rightarrow 6p ~ ^2P_{3/2}$  & 6.345(5)$^a$ & $6.4890(50)^b$ &  $-93.01(15)$ \\
$\rightarrow 7p ~ ^2P_{1/2}$  & 0.27810(45)$^c$ &  10.2915(100) & $-37.22(10)$  \\
$\rightarrow 7p ~ ^2P_{3/2}$  & 0.57417(57)$^c$ & $-14.2703(120)$ & $-101.53(18)$ \\
$\rightarrow 8p ~ ^2P_{1/2}$  & 0.0824(10) & $0.9623(20)$ & $-0.52(1)$ \\
$\rightarrow 8p ~ ^2P_{3/2}$  & 0.2294(15) & $-1.7115(20)$ & $-2.54(2)$ \\
$\rightarrow 9p ~ ^2P_{1/2}$  & $-0.0424(15)$ & $-0.3896(15)$ & $-0.08(1)$ \\
$\rightarrow 9p ~ ^2P_{3/2}$  & 0.1268(10) & $-0.7388(20)$ & $-0.50(1)$ \\
\hline \\
Core &     &    & 0.1999(50)  \\
$n>9$ &    &    & $-0.8547(500)$ \\
\hline \\
Total  &   &    &  $-268.65(27)$    \\
  \end{tabular}
   \end{ruledtabular}
   References:  $^a$\cite{gregoire}; $^b$\cite{Toh};  $^c$\cite{damitz}. \\
 \label{tab7}
\end{table}

\subsection{Vector polarizability}

An accurate determination of the vector ($\beta$) dipole polarizability of the $6s ~ ^2S_{1/2} - 7s ~ ^2S_{1/2}$ transition in $^{133}$Cs is 
imperative so that it can be combined with the measured value of $Im(E1_{PV}/\beta)$ and our high accuracy calculation of $E1_{PV}$   
to extract $Q_W^{\rm at}$. A very precise measurement of $\alpha/\beta=-9.905(11)$ has been reported by Cho et al. \cite{cho}, where 
$\alpha$ is the scalar dipole polarizability of the transition. The $\alpha$ value for the transition $|\Psi_i \rangle \rightarrow |\Psi_f\rangle$ 
can be expressed as \cite{blundell} 
\begin{eqnarray}
  \alpha &=& \sum_{k} \frac{ \langle \Psi_f^{(0)} | D | \Psi_k^{(0)} \rangle \langle \Psi_k^{(0)} | D | \Psi_i^{(0)} \rangle}
  {\sqrt{\langle \Psi_f^{(0)} | \Psi_f^{(0)} \rangle \langle \Psi_i^{(0)} | \Psi_i^{(0)} \rangle}} \nonumber \\
  && \ \ \ \ \ \ \ \ \ \ \ \ \times \left [ \frac{1}{E_f^{(0)}-E_k^{(0)}}  + \frac{1}{E_i^{(0)}-E_k^{(0)}}  \right ].  
\label{alphaeq}
\end{eqnarray}
As in the case of $E1_{PV}$, the contributions to $\alpha$ come from the ``Core'', ``Main'' and ``Tail'' regions. We have included the E1 matrix 
elements up to the $9P$ states in this estimation. Most of these matrix elements were calculated in the present work using the RCCSDT method, except a few for 
which very accurate experimental data are available  \cite{gregoire,Toh,damitz}. We have also used measured values of the energies in our calculations.
 The contributions from the ``Core'' and ``Tail'' were estimated to be small using the RMP(2) method. The individual contributions from ``Main'' that 
 come from the low-lying intermediate states, ``Core'' and ``Tail'' are given in Table \ref{tab7}. The matrix elements used from different works are presented in the same table. As can be seen from the table, the maximum contribution to $\alpha$ of the $6s ~ ^2S_{1/2} - 7s ~ ^2S_{1/2}$ transition in $^{133}$Cs 
 comes from the $7p ~ ^2P_{3/2}$ state followed by the $6p ~ ^2P_{3/2}$ state. The contributions from the $8P$ state onwards are found to be small. 
 Our final value is $\alpha=-268.65(27)ea_0^3$. Another recent study has found this value to be $-268.82(30)ea_0^3$ \cite{toh1}, where 
 contributions from many matrix elements were 
included explicitly by analyzing them from the literature. They had estimated  the ``Core'' and ``Tail'' contributions using the DHF method, 
whereas we have done so using the RMP(2) method. Nonetheless, we find very good agreement between both the results. By combining our value for 
 $\alpha$  with the measured ratio of $\alpha/\beta=-9.905(11)$ \cite{cho}, we obtain the vector polarizability for this transition as $\beta=27.12(4) \ ea_0^3$.
The accuracy of this quantity is about 0.15\%; even better than the accuracy of our calculated $E1_{PV}$ for the above transition. In Ref. 
\cite{toh1}, a summary of the results for $\beta$ have been presented, the variation in  these values covers a wide range. Our result is in agreement 
with all those values, but with a precision similar to the most accurate one \cite{toh1}.

\subsection{Inferred $Q_W$ value and its implications}

Combining our results of $E1_{PV}$ and $\beta$ with the precisely measured $Im(E1_{PV}/\beta)=1.5935(56)$ mV/cm \cite{wood}, where $Im$ 
means imaginary part, for the $6s ~ ^2S_{1/2} - 7s ~ ^2S_{1/2}$ transition in $^{133}$Cs, we get $Q_W^{\rm at}=-73.54(26)_{ex}(22)_{th}$. 
After accounting for the nuclear skin effect \cite{sil}, we get
\begin{eqnarray}
 Q_W &=& Q_W^{\rm at} + \Delta Q_W^{N-P} \nonumber \\ &=& -73.54(26)_{ex}(22)_{th} + 0.064 \nonumber \\ &=& -73.48(26)_{ex}(23)_{th}.
\end{eqnarray}
This results in the difference between the value of $Q_W$ obtained from our calculation and the SM value $Q_W^{\text{SM}}=-73.23(1)$ \cite{tanabashi} 
as $\Delta Q_W \equiv Q_W -Q_W^{\text{SM}}=-0.25(34)$ at 1$\sigma$ level. 

\begin{figure}[t]
\includegraphics[width=8.5cm,height=6cm]{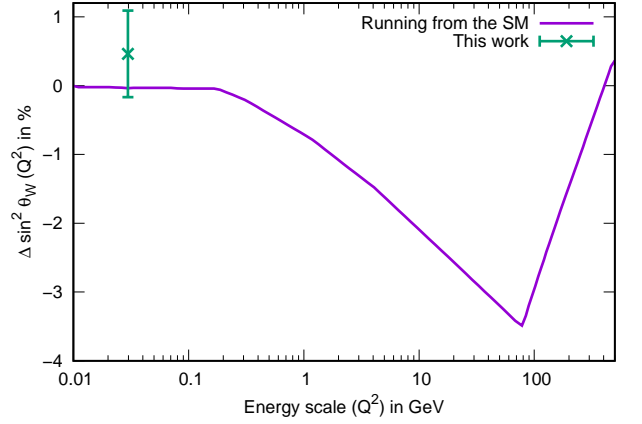}
\caption{\label{fig1} Plot demonstrating deviation of $\Delta \sin \theta_W (Q^2)$ (in percentage) in the SM from 
$\sin^2 \theta_W(m_{Z_0})_{\overline{\rm MS}} = 0.23124(6)$ with energy scale ($Q$) in GeV. The value obtained using the present APV study in 
$^{133}$Cs is shown at $Q^2=30$ MeV, which shows good agreement with the SM.} 
\end{figure}

From the relation $Q_W= -N+Z(1-4\sin^2\theta_W)$, we can derive as
\begin{eqnarray}
Q_W & \approx & -N+Z [ 1-4 (\sin^2\theta_W^{\rm SM} +\Delta \sin^2 \theta_W) ] \nonumber \\
    & = & Q_W^{\rm SM} -4 Z \Delta \sin^2 \theta_W . \nonumber \\
\Rightarrow \Delta Q_W & \approx & -4 Z \Delta \sin^2 \theta_W .
\end{eqnarray}
This gives change in $\sin^2\theta_W$ as $\Delta (\sin^2 \theta_W)=0.0011(15)$. Accounting for this correction along with its SM value 
$\sin^2 \theta_W^{SM}=0.23857(5)$ at the zero momentum transfer in the $\overline{\text{MS}}$ scheme \cite{tanabashi}, we get a new value for 
$\sin^2 \theta_W (0)=0.23967(150)$. In Fig. \ref{fig1}, we plot deviation in running of $\Delta \sin^2 \theta_W(Q^2)$ with respect to 
$\sin^2 \theta_W(m_{Z_0})_{\overline{\rm MS}}$ from the SM and the deviation obtained in this work at $Q^2=30$ MeV corresponding 
to the experiment on $^{133}$Cs \cite{wood,Milstein}. It can be seen that the $\Delta \sin^2 \theta_W$ value obtained from the present
study agrees quite well with the SM. 

From the above $\Delta Q_W$ value, we constrain the isospin conserving parameter $S \simeq 0.31(43)$ after neglecting the contribution 
from the isospin breaking parameter $T$ from the relation $\Delta Q_W \approx -0.8S-0.007T$ \cite{marciano}. Furthermore, in the SO(10) model
 \cite{marciano}
\begin{eqnarray} 
 \Delta Q_W \approx 0.4 \times (2N+Z) \frac{M_W^2}{M_{Z_x}^2},
\end{eqnarray}
we get a lower limit $M_{Z_x} > 961$ GeV/$c^2$ compared to 3.5 TeV/$c^2$ from the observation using the ATLAS detector \cite{herrero}. Furthermore, 
$\Delta Q_W$ can be expressed in the dark photon model characterized by $U(1)_d$ gauge symmetry as \cite{hooman}
\begin{eqnarray}
 \Delta Q_W = 220 \left ( \frac{\varepsilon}{\varepsilon_z}\right ) \sin \theta_W  \cos \theta_W  \delta^2 - Q_W^{SM} \delta^2,
\end{eqnarray}
where $\varepsilon$ is a dimensionless parameter, $\varepsilon_z= \delta M_{Z_d}/M_{Z_0}$, and $\delta$ is a model dependent quantity. 
Substituting the aforementioned SM values, we get     
\begin{eqnarray}
  \left [ 1.28(1) \left ( \frac{\varepsilon}{\varepsilon_z}\right ) - 1 \right ] \delta^2 \simeq 0.0034(46).
\end{eqnarray}  

Using the effective field theory, suggested in Ref. \cite{arcadi}, we obtain  
\begin{eqnarray}
 f_{Vq}^{eff}/\Lambda^2 \simeq -26(35) \times 10^{-10} {\rm GeV}^{-2} .
\end{eqnarray}
Similarly, in terms of the nucleon-electron V-A couplings, defined in Ref. \cite{tanabashi}, it yields
$55g^{ep}+78g^{en}=36.82(25)$.

\section{Summary}

We have revisited the calculation of electric dipole amplitude due to the nuclear spin independent neutral weak interaction for the 
$6s ~ ^2S_{1/2} - 7s ~ ^2S_{1/2}$ transition in $^{133}$Cs by employing the relativistic coupled-cluster theory. In our approach, we solve an 
inhomogeneous equation to obtain the first-order perturbed wave function due to the weak interaction in order to account for the
correlation effects of the electrons from the occupied, valence and virtual orbitals on an equal footing. This resolves the
large discrepancy, including sign, for the core electron correlation contribution to the above amplitude between the two latest high accuracy
calculations. Moreover, it includes contributions from correlation effects due to the double core-polarization, the Breit interaction and lower-order quantum electrodynamics effects by the same method used 
to incorporate contributions from the Dirac-Coulomb atomic Hamiltonian. Relevant spectroscopic properties have been evaluated at different levels of 
many-body approximations and the role of electron correlation effects arising from higher-level particle-hole excitations, in particular the
triple excitations, have been demonstrated to be non-negligible. By analyzing the differences between these calculated results and their respective 
high-precision experimental values, the accuracy of the above electric dipole transition amplitude is estimated and found to be of the order of 0.3\%. This 
is slightly better than the reported accuracy of the corresponding measurement. We have determined the vector polarizability of the above
transition with an accuracy of 0.15\% . Combining all our calculated values with the measurement, we have obtained the nuclear weak charge
$Q_W=-73.48(26)_{ex}(23)_{th}$ for $^{133}$Cs , which differs from the Standard Model value by $-0.25(34)$. By considering certain extensions of
the Standard Model of current interest, we have discussed the salient implications of this discrepancy in the nuclear weak charge for possible new physics. Our findings are in agreement with the Standard Model predictions. 

\section*{Acknowledgement}

We are grateful to Professors Jens Erler, Hubert Spiesberger and Mikahil Gorshteyn for giving us an opportunity to present this work 
at MITP Virtual Workshop on ``Parity Violation and Related Topics''. Computations reported in this work were performed using the PRL Vikram-100 
HPC cluster.


\begin{thebibliography}{99}
\bibitem{Breit}
G. Breit, Phys. Rev. {\bf 34}, 553 (1929); Phys. Rev. {\bf 39}, 616 (1932).
\bibitem{Griffiths}
D. J. Griffiths, {\it Introduction to Quantum Mechanics}, Prentice Hall Inc., Upper Saddle River, New Jersey 07458 (1995).
\bibitem{Bouchiat}
M.-A. Bouchiat and C. Bouchiat, Phys. Lett. B {\bf 48}, 111 (1974); J. Phys. (France) {\bf 35}, 899 (1974).
\bibitem{Ginges1}
J. S. M. Ginges and V. V. Flambaum, Phys. Rep. {\bf 397}, 63 (2004).
\bibitem{Cahn}
R. N. Cahn and G. L. Kane, Phys. Lett. B {\bf 71}, 348 (1977).
\bibitem{Marciano00}
W. J. Marciano and A. I. Sanda, Phys. Rev. D {\bf 17}, 3055 (1978).
\bibitem{Commins}
E. D. Commins, Phys. Scr. {\bf T46}, 92 (1993). 
\bibitem{Meekhof}
D. M. Meekhof, P. A. Vetter, P. K. Majumder, S. K. Lamoreaux, and E. N. Fortson, Phys. Rev. A {\bf 52}, 1895 (1995).
\bibitem{wood}
C. S. Wood, S. C. Bennet, D. Cho, B. P. Masterson, J. L. Roberts, C. E. Tanner, and C. E. Wieman, Science {\bf 275}, 1759 (1997).
\bibitem{bouchiat1}
M.-A. Bouchiat and C. Bouchiat, Rep. Prog. Phys. {\bf 60}, 1351 (1997).
\bibitem{Tsigutkin}
K. Tsigutkin, D. Dounas-Frazer, A. Family, J. E. Stalnaker, V. V. Yashchuk, and D. Budker, Phys. Rev. Lett. {\bf 103}, 071601 (2009).
\bibitem{marciano}
W. J. Marciano and J. L. Rosner, Phys. Rev. Lett. {\bf 65}, 2963 (1990); {\bf 68}, 898(E) (1992).
\bibitem{erler}
J. Erler and M. J. Ramsey-Musolf, Phys. Rev. D {\bf 72}, 073003 (2005).
\bibitem{kumar}
K. S. Kumar, S. Mantry, W. J. Marciano, and P. A. Soude, Ann. Review Nuc. Part. Science {\bf 63}, 237 (2013).
\bibitem{Cirigliano}
V. Cirigliano and M. J. Ramsey-Musolf, Prog. Part. Nuc. Phys. {\bf 71}, 2 (2013).
\bibitem{derevianko1}
M. S. Safronova, D. Budker, D. DeMille, D. F. J. Kimball, A. Derevianko and C. W. Clark, Rev. of Mod. Phys. {\bf 90}, 025008 (2018).
\bibitem{choi}
J. Choi and D. S. Elliott, Phys. Rev. A {\bf 93}, 023432 (2016).
\bibitem{anders}
A. Kastberg, T. Aoki, B. K. Sahoo, Y. Sakemi and B. P. Das, Phys. Rev. A {\bf 100}, 050101(R) (2019).
\bibitem{dzuba1}
V. A. Dzuba, J. C. Berengut, V. V. Flambaum, and B. Roberts, Phys. Rev. Lett. {\bf 109}, 203003 (2012).
\bibitem{tanabashi}
M. Tanabashi {\it et al.} (Particle Data Group), Phys. Rev. D {\bf 98}, 030001 (2018).
\bibitem{hooman}
H. Davoudiasl, H. -S. Lee and W. J. Marciano, Phys Rev. D {\bf 85}, 115019 (2012).
\bibitem{marciano1}
W. J. Marciano, AIP Conference Proceedings {\bf 1563}, 90 (2013).
\bibitem{davoudias}
H. Davoudiasl, H.-S. Lee and W. J. Marciano, Phys. Rev. D {\bf 92}, 055005 (2015).
\bibitem{arcadi}
G. Arcadi, M. Lindner, J. Martins and F. S. Queiroz, arXiv:1906.04755 (2019).
\bibitem{Banerjee}
D. Banerjee et al. (NA64 Collaboration), Phys. Rev. Lett. {\bf 118}, 011802 (2017); Phys. Rev. Lett. {\bf 120}, 231802 (2018); 
Phys. Rev. Lett. {\bf 123}, 121801 (2019).
\bibitem{dzuba2}
V. A. Dzuba, V. V. Flambaum, and O. P. Sushkov, Phys. Lett. A {\bf 141}, 147 (1989).
\bibitem{blundell}
S. A. Blundell, W. R. Johnson, and J. Sapirstein, Phys. Rev. Lett. {\bf 65}, 1411 (1990); Phys. Rev. D {\bf 45}, 1602 (1992).
\bibitem{Derevianko3}
A. Derevianko, Phys. Rev. Lett. {\bf 85}, 1618 (2000); Phys. Rev. A {\bf 65}, 012106 (2001).
\bibitem{kozlov}
M. G. Kozlov, S. G. Porsev, and I. I. Tupitsyn, Phys. Rev. Lett. {\bf 86}, 3260 (2001).
\bibitem{shabaev}
V. M. Shabaev, K. Pachucki, I. I. Tupitsyn, and V. A. Yerokhin, Phys. Rev. Lett. {\bf 94}, 213002 (2005).
\bibitem{porsev1}
S. G. Porsev, K. Beloy, and A. Derevianko, Phys. Rev. Lett. {\bf 102}, 181601 (2009); Phys. Rev. D {\bf 82}, 036008 (2010).
\bibitem{roberts2}
B. M. Roberts, V. A. Dzuba and V. V. Flambaum, Phys. Rev. A {\bf 88}, 042507 (2013).
\bibitem{derevianko2}
C. Wieman and A. Derevianko, arXiv:1904.00281
\bibitem{sil}
T. Sil, M. Centelles, X. Vias and J. Piekarewicz, Phys. Rev. C {\bf 71}, 045502 (2005).
\bibitem{bishop}
R. F. Bishop, {\it Microscopic Quantum Many-Body Theories and their Applications}, Lecture Series in Physics, pg. 1, Springer Publication, Berlin (1998).
\bibitem{bartlett}
I. Shavitt and R. J. Bartlett, {\it Many-body methods in Chemistry and Physics}, Cambidge University Press, Cambridge, UK (2009).
\bibitem{crawford}
T. D. Crawford and H. F. Schaefer, Rev. Comp. Chem. {\bf 14}, 33 (2000).
\bibitem{Dirac}
P. A. M. Dirac, {\it Principles of Quantum Mechanics}. International Series of Monographs on Physics (4th ed.), Oxford University Press (1982) [1958]. 
\bibitem{breit}
G. Breit, Phys. Rev. {\bf 34}, 553 (1929).
\bibitem{Uehl}
E. A. Uehling, Phys. Rev. 48 {\bf 55}, (1935).
\bibitem{Wichmann}
E. H. Wichmann and N. H. Kroll, Phys. Rev. {\bf 101}, 843 (1956).
\bibitem{Flambaum}
V. V. Flambaum and J. S. M. Ginges, Phys. Rev. A {\bf 72}, 052115 (2005).
\bibitem{Ginges}
J. S. M. Ginges and J. C. Berengut, Phys. Rev. A {\bf 93}, 052509 (2016).
\bibitem{Hofstadter}
R. Hofstadter, H. R. Fechter and J. A. McIntyre, Physical Review {\bf 92}, 978 (1953).
\bibitem{angeli}
I. Angeli, Atomic Data Nuc. Data Tables {\bf 87}, 185 (2004).
\bibitem{lindgren}
I. Lindgren and J. Morrison, {\it Atomic Many-Body Theory}, ed. by J. P. Toennies,  Springer-Verlag (Berlin) 1982.
\bibitem{mukherjee}
D. Mukherjee and S. Pal, Adv. Quant. Chem. {\bf 20}, 281 (1989).
\bibitem{bksahoo1}
B. K. Sahoo, D. K. Nandy, B. P. Das and Y. Sakemi, Phys. Rev. A {\bf 91}, 042507 (2015).
\bibitem{bksahoo2}
B. K. Sahoo and B. P. Das, Phys. Rev. A {\bf 92}, 052511 (2015).
\bibitem{bksahoo3}
B. K. Sahoo, Phys. Rev. A {\bf 93}, 022503 (2016).
\bibitem{bijaya1}
B. K. Sahoo, R. K. Chaudhuri, B. P. Das and D. Mukherjee, Phys. Rev. Lett. {\bf 96}, 163003 (2006).
\bibitem{bijaya2}
L. Wansbeek, B. K. Sahoo, R. G. E. Timmermans, K. Jungmann, B. P. Das, and D. Mukherjee, Phys. Rev. A {\bf 78}, 050501(R) (2008).
\bibitem{bijaya3}
B. K. Sahoo and B. P. Das, Phys. Rev. A {\bf 84}, 010502(R) (2011).
\bibitem{nandy}
D. K. Nandy and B. K. Sahoo, Phys. Rev. A {\bf 90}, 050503(R) (2014).
\bibitem{shabaev1}
V. M. Shabaev, M. Tomaselli, T. K\"uhl, A. N. Artemyev, and V. A. Yerokhin, Phys. Rev. A {\bf 56}, 252 (1997).
\bibitem{shabaev2}
A. V. Volotka, D. A. Glazov, I. I. Tupitsyn, N. S. Oreshkina, G. Plunien, and V. M. Shabaev, Phys. Rev. A {\bf 78}, 062507 (2008).
\bibitem{Boys}
S. F. Boys, Proc. R. Soc. Lond. A. {\bf 200}, 542 (1950).
\bibitem{grasp2}
K. G. Dyall, I. P. Grant, C. T. Johnson, F. A. Parpia and E. P. Plummer, Comp. Phys. Comm. {\bf 55}, 425 (1989).
\bibitem{bennett}
S. C. Bennett and C. E. Wieman, Phys. Rev. Lett. {\bf 82}, 2484 (1999).
\bibitem{Roberts}
B. M. Roberts, V. A. Dzuba, and V. V. Flambaum, Phys. Rev. A {\bf 87}, 054502 (2013).
\bibitem{nist}
A. Kramida, Yu. Ralchenko, J. Reader, and NIST ASD Team (2018), {\it NIST Atomic Spectra Database} (ver. 5.6.1), 
 National Institute of Standards and Technology, Gaithersburg, MD.
\bibitem{arimondo}
E. Arimondo, M. Inguscio, and P. Violino, Rev. Mod. Phys. {\bf 49}, 31 (1977).
\bibitem{dipankar}
D. Das and V. Natarajan, J. Phys. B {\bf 39}, 2013 (2006).
\bibitem{yang}
Guang Yang, Jie Wang, Baodong Yang, and Junmin Wang, Laser Phys. Letts. {\bf 13}, 085702 (2016).
\bibitem{williams}
W. D. Williams, M. T. Herd and W. B. Hawkins, Laser Phys. Letts. {\bf 15}, 095702 (2018).
\bibitem{happer}
W. Happer, {\it Atomic Physics 4}, eds. G. zu Putlitz, E. W. Weber, and A. Winnacker, (Plenum Press, New York) pp. 651-682 (1974).
\bibitem{gregoire}
M. D. Gregoire, I. Hromada, W. F. Holmgren, R. Trubko, and A. D. Cronin, Phys. Rev. A {\bf 92}, 052513 (2015).
\bibitem{damitz}
A. Damitz, G. Toh, E. Putney, C. E. Tanner and D. S. Elliott, Phys. Rev. A {\bf 99}, 062510 (2019).
\bibitem{Toh}
G. Toh, A. Damitz, N. Glotzbach, J. Quirk, I. C. Stevenson, J. Choi, M. S. Safronova and D. S. Elliot, Phys. Rev. A {\bf 99}, 032504 (2019).
\bibitem{Bennett1}
S. C. Bennett, J. L. Roberts, and C. E. Wieman, Phys. Rev. A {\bf 59}, R16 (1999).
\bibitem{Young}
L. Young, W. T. Hill, III, S. J. Sibener, S. D. Price, C. E. Tanner, C. E. Wieman and S. R. Leone, Phys. Rev. A {\bf 50}, 2174 (1994).
\bibitem{Vasilyev}
A. A. Vasilyev, I. M. Savukov, M. S. Safronova, and H. G. Berry, Phys. Rev. A {\bf 66}, 020101 (2002).
\bibitem{Rafac}
R. J. Rafac, C. E. Tanner, A. E. Livingston, and H. G. Berry, Phys. Rev. A {\bf 60}, 3648 (1999).
\bibitem{Patternson}
B. M. Patterson, J. F. Sell, T. Ehrenreich, M. A. Gearba, G. M. Brooke, J. Scoville, and R. J. Knize, Phys. Rev. A {\bf 91}, 012506 (2015).
\bibitem{Antypas}
D. Antypas and D. S. Elliott, Phys. Rev. {\bf 88}, 052516 (2013).
\bibitem{Bouch}
M. -A. Bouchiat, J. Guena, and L. Pottier, J. Phys. (Paris), Lett. {\bf 45}, 523 (1984).
\bibitem{Milstein}
A. I. Milstein, O. P. Sushkov, and I. S. Terekhov, Phys. Rev. Lett. {\bf 89}, 283003 (2002).
\bibitem{Dzubab}
V. A. Dzuba, V. V. Flambaum, and J. S. M. Ginges, Phys. Rev. D {\bf 66}, 076013 (2002).
\bibitem{dzuba0}
V. A. Dzuba, V. V. Flambaum, and J. S. M. Ginges, Phys. Rev. D {\bf 66}, 076013 (2002).
\bibitem{toh1}
G. Toh, A. Damitz, C. E. Tanner, W. R. Johnson and D. S. Elliott, Phys. Rev. Lett. {\bf 123}, 073002 (2019).
\bibitem{cho}
D. Cho, C. S. Wood, S. C. Bennett, J. L. Roberts, and C. E. Wieman, Phys. Rev. A {\bf 55}, 1007 (1997).
\bibitem{herrero}
Tomas Davidek and Luca Fiorini, Front. Phys. {\bf 8}, 149 (2020). 

\end{thebibliography}
\end{document}